\documentclass[journal, compsoc]{IEEEtran}
\usepackage{amsmath,amsfonts}
\usepackage{algorithmic}
\usepackage{algorithm}
\usepackage{array}
\usepackage{mathtools}
\usepackage{caption}
\usepackage{subcaption}
\usepackage{textcomp}
\usepackage{stfloats}
\usepackage{url}
\usepackage{verbatim}
\usepackage{graphicx}
\usepackage{xcolor}
\usepackage{cite}
\usepackage{hyperref}
\usepackage[capitalise]{cleveref}
\usepackage{orcidlink}
\graphicspath{{figs/}{figures/}{pictures/}{images/}{./}} 

\DeclareMathOperator{\acosh}{acosh}
\DeclareMathOperator{\avg}{avg}
\DeclareMathOperator{\KL}{KL}
\DeclareMathOperator{\TP}{TP}
\DeclareMathOperator{\PR}{PR}
\DeclareMathOperator{\RC}{RC}
\DeclareMathOperator{\midd}{mid}
\DeclareMathOperator{\proj}{proj}
\DeclareMathOperator{\std}{std}

\begin{document}

\title{Accelerating hyperbolic t-SNE}

\author{
    Martin Skrodzki\orcidlink{0000-0002-8126-0511},
    Hunter van Geffen, 
    Nicolas~F.~Chaves-de-Plaza\orcidlink{0000-0003-4971-3151},\\ 
    Thomas H\"ollt\orcidlink{0000-0001-8125-1650}, 
    Elmar Eisemann\orcidlink{0000-0003-4153-065X}, 
    and Klaus Hildebrandt\orcidlink{0000-0002-9196-3923}
    \IEEEcompsocitemizethanks{
        \IEEEcompsocthanksitem Corresponding author: Martin Skrodzki, mail@ms-math-computer.science. Shared first authorship with Hunter van Geffen.
        \IEEEcompsocthanksitem All authors are affiliated with TU Delft, The Netherlands.
    }
}

\markboth{IEEE Transactions on Visualization and Computer Graphics}{Skrodzki \MakeLowercase{\textit{et al.}}: Accelerating hyperbolic t-SNE}

\IEEEpubid{}

\IEEEcompsoctitleabstractindextext{
\begin{abstract}
    The need to understand the structure of hierarchical or high-dimensional data is present in a variety of fields.
    Hyperbolic spaces have proven to be an important tool for embedding computations and analysis tasks as their non-linear nature lends itself well to tree or graph data.
    Subsequently, they have also been used in the visualization of high-dimensional data, where they exhibit increased embedding performance.
    However, none of the existing dimensionality reduction methods for embedding into hyperbolic spaces scale well with the size of the input data.
    That is because the embeddings are computed via iterative optimization schemes and the computation cost of every iteration is quadratic in the size of the input.
    Furthermore, due to the non-linear nature of hyperbolic spaces, Euclidean acceleration structures cannot directly be translated to the hyperbolic setting.
    This paper introduces the first acceleration structure for hyperbolic embeddings, building upon a polar quadtree.
    We compare our approach with existing methods and demonstrate that it computes embeddings of similar quality in significantly less time.
    
    Implementation and scripts for the experiments can be found at \url{https://graphics.tudelft.nl/accelerating-hyperbolic-tsne}.
  %
\end{abstract}

}


\maketitle

\section{Introduction}

The analysis of high-dimensional data is of major importance for a wide range of applications across many industry and research fields.
Dimensionality reduction is a key part of processing pipelines to visualize and analyze such data, which has recently been demonstrated in the application settings of sports~\cite{wu2021tacticflow}, literature search~\cite{narechania2021vitality}, machine learning~\cite{wang2021visual}, and e-commerce~\cite{zhang2022promotionlens}.
Effective embeddings of data points preserve structures in the data set, such that a visual inspection of the low-dimensional embedded data can help to gain insights into the structures of the high-dimensional data.
A widespread technique to create such embeddings is t-distributed stochastic neighbor embedding (t-SNE)~\cite{vanDerMaaten2008visualizing}.
It is popular because t-SNE preserves local neighborhoods particularly well when embedding the data, see \cref{sec:RelatedWork}.
Most dimensionality reduction algorithms traditionally embed data into flat, Euclidean space.
This misses the opportunities provided by other embedding spaces, like negatively curved, hyperbolic spaces.

Hyperbolic spaces already find applications in the embedding of trees, graphs, and other hierarchical data.
For example, it is possible to embed trees into two-dimensional hyperbolic space with arbitrarily low distortion~\cite{sarkar2011low}.
By this property, previous works have successfully embedded social networks~\cite{verbeek2014metric} or the Internet~\cite{boguna2010sustaining} into hyperbolic space.
Furthermore, hyperbolic spaces exhibit a natural \emph{Focus+Context} view of the data~\cite{lamping1995focus+, miller2022browser}, which significantly increases information foraging~\cite{pirolli2001visual}.
Finally, it has recently been suggested that hyperbolic spaces are suitable for navigating higher-dimensional spaces directly~\cite{kopczynski2021navigating}.

Given the utility of hyperbolic spaces for the visualization of hierarchical data, several methods have been proposed to translate t-SNE to work in hyperbolic space~\cite{guo2022co,klimovskaia2020poincare,zhou2021hyperbolic}.
These adaptions have shown great potential, when used, for instance, in visualization, clustering, lineage detection, and pseudotime inference tasks~\cite{klimovskaia2020poincare}.
We will discuss these in detail in \cref{sec:HyperbolicEmbeddings}.
Although they all create useful embeddings of high-dimensional data in hyperbolic spaces, solving their respective optimization problems is costly compared to methods that embed into Euclidean space.

The long optimization run time in hyperbolic spaces is mostly because accelerations for the computation of Euclidean embeddings are not directly effective for embeddings in hyperbolic spaces. 
For Euclidean embeddings, state-of-the-art implementations make use of acceleration methods~\cite{vanDerMaaten2014accelerating,linderman2019fast,pezzotti2019gpgpu,vanDeRuit2021efficient}, which have been developed over the last years to speed up the processing, see \cref{sec:Acceleratingt-SNE}.
However, contrary to flat Euclidean spaces, hyperbolic spaces exhibit negative curvature. 
One consequence is that the circumference and area of a circle in a two-dimensional hyperbolic space grow exponentially with its radius, while they grow polynomially in Euclidean space ~\cite{krioukov2010hyperbolic}. 
On the one hand, these properties make hyperbolic spaces well-suited for the embedding of structures that also grow exponentially. On the other hand, it also leads to the lack of linear interpolation or averages in these non-linear, hyperbolic spaces.
Euclidean acceleration structures, including those listed above, rely on these properties and can thus not be translated directly for use in hyperbolic spaces.
For example, the Barnes-Hut scheme~\cite{vanDerMaaten2014accelerating} accelerates the optimization by building a quadtree on the embedding space.
Here, equal-sized quadrilaterals form the nodes of the tree and their midpoints act as accelerating proxies.
In hyperbolic spaces, no direct analog of such a tree can be built due to the exponential growth and non-linear properties of such spaces.
Instead, existing approaches for hyperbolic t-SNE turn to sampling the data in order to compute the embeddings within a reasonable time frame.
This limits the visualization of the data to only a limited portion of the input.

This paper introduces the first acceleration structure for hyperbolic embeddings. 
We use a polar quadtree~\cite{looz2015generating}, designed to operate in hyperbolic spaces.
However, we find that the data structure needs to be adjusted to the specific setting of embeddings by changing its build procedure to provide a reliable speed-up of the optimization.
Based on this modified data structure, we proceed to formulate an approximation of the cost function gradient used in the optimization of hyperbolic t-SNE embeddings.
By analyzing the gradients of current state-of-the-art approaches for hyperbolic embeddings~\cite{guo2022co,klimovskaia2020poincare,zhou2021hyperbolic}, we show that our acceleration technique can be adjusted to their respective needs.
Thus, it is a versatile building block for current and future hyperbolic embedding approaches.
Finally, we present several experiments to validate our findings and conclusions.
In summary, the contributions of this paper are:
\begin{itemize}
    \item the presentation of a polar quadtree data structure for hyperbolic embedding computations,
    \item a new splitting rule for the data structure that enhances performance for embedding computations,
    \item a fast approximation scheme of hyperbolic gradient descent iterations using the data structure, and
    \item an analysis of how to integrate this approximation into existing approaches for hyperbolic embeddings.
\end{itemize}


\section{Background}

In this section, we present the techniques and concepts that our method is built upon.
Specifically, we introduce t-SNE, its Barnes-Hut acceleration for Euclidean embeddings, and necessary concepts of hyperbolic spaces.
Finally, we present a hyperbolic data structure that was designed for fast random graph generation in hyperbolic space and that will serve as a basis for our acceleration of hyperbolic t-SNE.

\subsection{t-distributed Stochastic Neighbor Embedding}
\label{sec:t-SNE}

A widely used technique for non-linear dimensionality reduction is t-distributed Stochastic Neighbor Embedding (t-SNE), which creates a low-dimensional embedding of the data while aiming at preserving local neighborhoods of the high-dimensional data points~\cite{vanDerMaaten2008visualizing}.
This is achieved by interpreting the high-dimensional input~${\{\mathbf{x}_1,\ldots,\mathbf{x}_n\}\subseteq\mathbb{R}^d}$ as (conditional) probabilities by
\begin{align}
\label{equ:HighDimensionalProbability}
    p_{j|i} = \frac{\exp\left(-\left\|\mathbf{x}_i-\mathbf{x}_j\right\|^2/2\sigma_i\right)}{\sum_{k\neq i}\exp\left(-\left\|\mathbf{x}_i-\mathbf{x}_k\right\|^2/2\sigma_i^2\right)}, \quad 
    p_{ij} = \frac{p_{j|i}+p_{i|j}}{2},
\end{align}
where~$p_{i|i}=0$ and~$\sigma_i$ is the variance of the Gaussian centered on point~$\mathbf{x}_i$.
In practice,~$\sigma_i$ is chosen such that the perplexity of the probability distribution~$P_i$ equals a user-prescribed perplexity value.
On the low-dimensional embedding~${Q=\{\mathbf{y}_1,\ldots,\mathbf{y}_n\}\subseteq\mathbb{R}^{d'}}$, a corresponding probability distribution is given by
\begin{align}
\label{equ:LowDimensionalProbability}
    q_{ij}=\frac{\left(1+\left\|\mathbf{y}_i-\mathbf{y}_j\right\|^2\right)^{-1}}{\sum_{k\neq \ell}\left(1+\left\|\mathbf{y}_k - \mathbf{y}_\ell\right\|^2\right)^{-1}}.
\end{align}
To compute the positions~$\mathbf{y}_i$ of the low-dimensional embedding, t-SNE starts with an initial embedding obtained by principal component analysis (PCA)~\cite{kobak2021initialization} and then alters the embedding by gradient-descent optimization of the Kullback-Leibler divergence between the high- and the low-dimensional probability distribution, which is given by
\begin{align}
\label{equ:KLDivergence}
    C = \KL(P||Q) = \sum_i\sum_j p_{ij}\log \frac{p_{ij}}{q_{ij}},
\end{align}
with the gradient
\begin{align}
\label{equ:KLGradient}
    \frac{\delta C}{\delta \mathbf{y}_i} = 4\sum_{j\neq i}(p_{ij}-q_{ij})\left(1+\left\|\mathbf{y}_i-\mathbf{y}_j\right\|^2\right)^{-1}(\mathbf{y}_i-\mathbf{y}_j).
\end{align}
The naive implementation of t-SNE has a run time of~$\mathcal{O}(n^2)$ as evaluating the gradient takes quadratic time in the number of input points.
This is clear from rewriting $\delta C/\delta \mathbf{y}_i$ as
\begin{align}
\label{equ:KLGradientSplit}
    \frac{\delta C}{\delta \mathbf{y}_i} = 4\left(\sum_{j\neq i} p_{ij}q_{ij}Z(\mathbf{y}_i-\mathbf{y}_j) - \sum_{j\neq i}q_{ij}^2Z(\mathbf{y}_i-\mathbf{y}_j)\right),
\end{align}
where~$Z=\sum_{k\neq\ell}\left(1+\left\|\mathbf{y}_k - \mathbf{y}_\ell\right\|^2\right)^{-1}$.
The first sum can be computed efficiently, if the probability distribution~$P$ is sparse~\cite{vanDerMaaten2014accelerating}, that is if the Gaussians in \cref{equ:HighDimensionalProbability} are truncated.
However, the second sum requires~$\mathcal{O}(n^2)$ operations.

\begin{figure}
    \centering
    \def\svgwidth{1.\linewidth}
    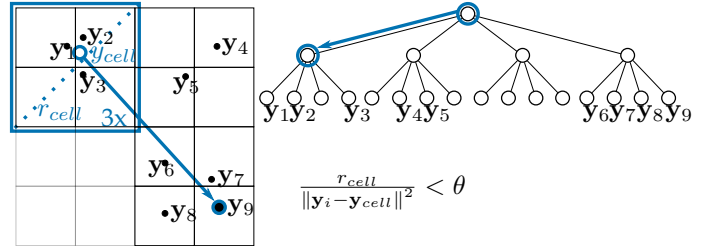
    \caption{
        The Barnes-Hut data structure, showing the quadtree and the hierarchy. 
        The influence of the points~$\mathbf{y}_1,\mathbf{y}_2,\mathbf{y}_3$ on point~$\mathbf{y}_9$ is approximated using their midpoint~$\mathbf{y}_{cell}$ and the cell diagonal~$r_{cell}$.
        Modified from~\cite{vanDerMaaten2014accelerating}.
    }
    \label{fig:BarnesHutApproximation}
\end{figure}

\subsection{Barnes-Hut Acceleration Structure for t-SNE}
\label{sec:BarnesHutAcceleration}

Several accelerating methods have been proposed to speed up the gradient computation.
A method inspired by $n$-body simulation is to build a quadtree data structure, alternatively called a Barnes-Hut tree, on the embedding points~\cite{vanDerMaaten2014accelerating}.
This hierarchical data structure enables the approximation of the second sum of \cref{equ:KLGradientSplit}.
It does so by grouping points~$\mathbf{y}_j$ far away from the query point~$\mathbf{y}_i$ on a higher level of the quadtree hierarchy and using a summary of the cell instead of the individual points.
That is, when evaluating the gradient for an embedding point~$\mathbf{y}_i$, we traverse the quadtree structure.
At every cell, we evaluate whether
\begin{align}
\label{equ:CellApproximationCriterion}
    \frac{r_{\text{cell}}}{\left\|\mathbf{y}_i-\mathbf{y}_{\text{cell}}\right\|} < \theta
\end{align}
holds, where~$r_{\text{cell}}$ is the length of the diagonal of the cell,~$\mathbf{y}_{\text{cell}}$ denotes the arithmetic midpoint of all points stored in the cell, and~$\theta$ is a user-given parameter to steer the approximation.
Typically,~$\theta$ is set somewhere between~$0.2$ to~$0.8$~\cite{vanDerMaaten2014accelerating}.
If \cref{equ:CellApproximationCriterion} holds, we do not further traverse the hierarchy but instead utilize the midpoint~$\mathbf{y}_{\text{cell}}$, weighted by the number of embedding points represented by the cell, in the evaluation of the gradient.
See \cref{fig:BarnesHutApproximation} for an illustration of this procedure.

\subsection{Hyperbolic Space and the Poincar\'e Disk Model}

\begin{figure}
    \centering
    \includegraphics[width=.35\linewidth]{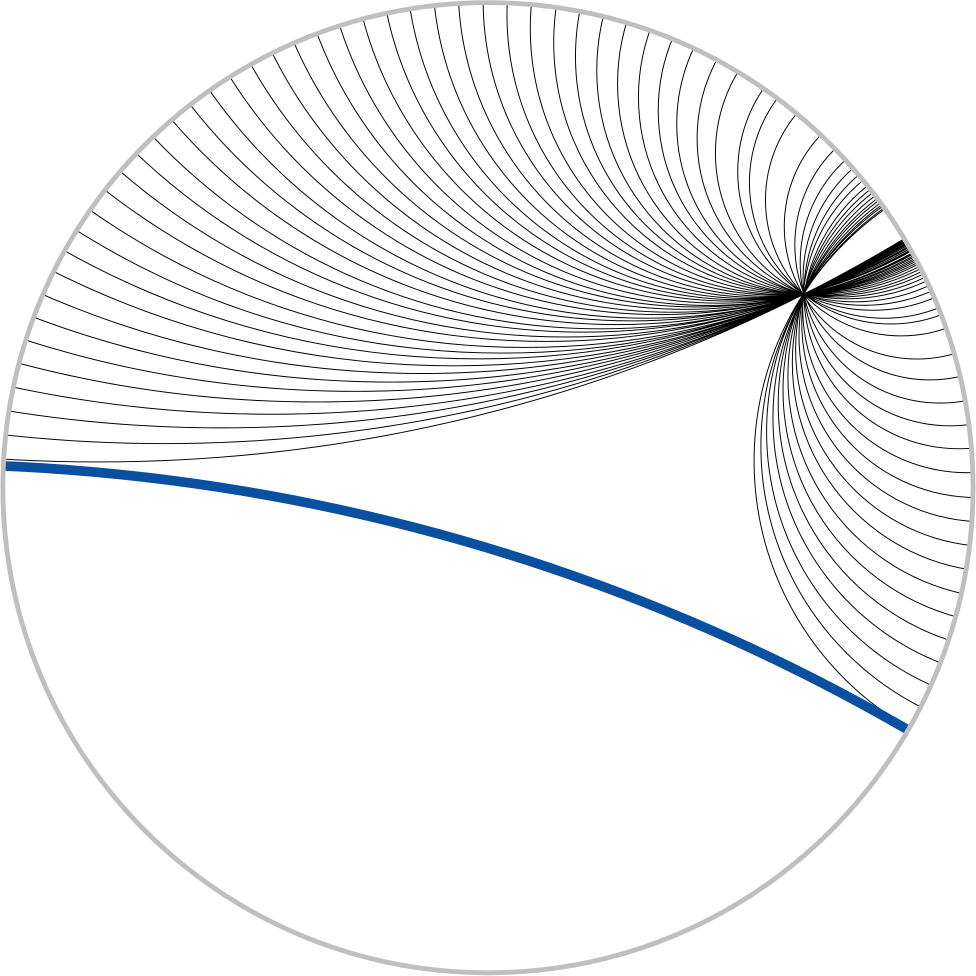}
    \hspace{.025\linewidth}
    \includegraphics[width=.35\linewidth]{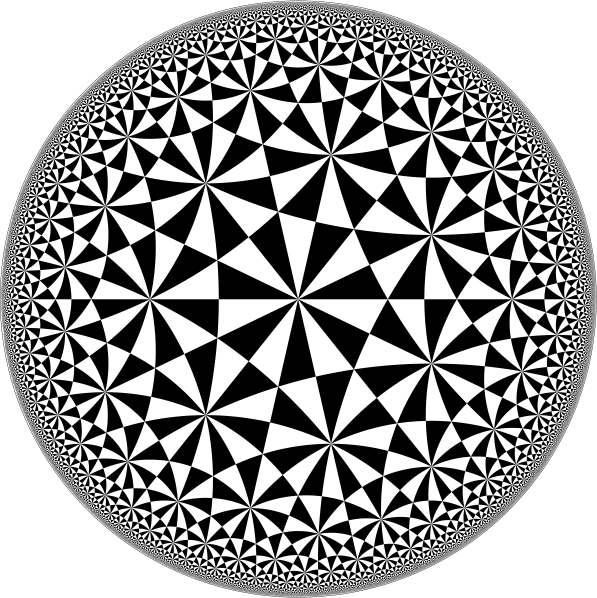}
    \caption{The Poincar\'e disk model of hyperbolic space with a blue straight line in the left that appears curved and has infinitely many parallels to it. On the right a regular tiling illustrating how tiles visually shrink towards the edge of the disk, while keeping their area within hyperbolic space.}
    \label{fig:PoincareDisk}
\end{figure}

As this work aims at embeddings in hyperbolic space, we will recall several important notions.
Working with hyperbolic space calls for choosing an appropriate model to work with, for example, the Poincar\'e disk model, the Lorentz hyperboloid model, or the Klein model~\cite[Sec.~7]{cannon1997hyperbolic}.
All these models are compatible with each other and translation from one model to another is not costly.
Our embeddings will be placed in the Poincar\'e disk, see \cref{fig:PoincareDisk}.
This is a suitable model because it maps the entire two-dimensional hyperbolic space to a finite disk. 
Furthermore, it has the advantage of being conformal, which helps in splitting the space into a hierarchy.
However, we will use the Klein model for the computation of the Einstein midpoint, see the discussion after \cref{equ:EinsteinMidpoint}.
Other embedding approaches have turned to the Lorentz model, because of better numerical precision~\cite{klimovskaia2020poincare}.
However, since we will build our acceleration structure directly on the Poincar\'e  disk, that is, on the embedding space, we obtain satisfactory results without translating to the Lorentz model.

Formally, the Poincar\'e disk model is the space~${\mathbb{D}=\{\mathbf{y}\in\mathbb{R}^2:\left\|\mathbf{y}\right\|<1\}}$ equipped with the metric
\begin{align}
\label{equ:PoincareMetric}
    g_\mathbf{y}^\mathbb{D} =\lambda_\mathbf{y}^2g^E, && \text{where} && \lambda_\mathbf{y} = \frac{2}{1-\left\|\mathbf{y}\right\|^2},
\end{align}
with~$g^E$ the standard scalar product of~$\mathbb{R}^2$ and~$\left\|.\right\|$ the standard norm of~$\mathbb{R}^2$, see~\cite[Eq.~(1)]{ganea2018hyperbolic}.
The hyperbolic distance $d^\mathcal{H}(\mathbf{y}_i,\mathbf{y}_j)$ between two points~$\mathbf{y}_i$ and~$\mathbf{y}_j$ in the Poincar\'e model is then given by~\cite[Eq.~(2)]{ganea2018hyperbolic}.
\begin{align}
\label{equ:PoincareDistance}
    \begin{split}
    d^\mathcal{H}_{ij} \coloneqq \cosh^{-1}\left(1+2\frac{\left\|\mathbf{y}_i-\mathbf{y}_j\right\|^2}{\left(1-\left\|\mathbf{y}_i\right\|^2\right)\left(1-\left\|\mathbf{y}_j\right\|^2\right)}\right).
    \end{split}
\end{align}

\subsection{Polar quadtree}
\label{sec:PolarQuadtree}

\begin{figure}
    \centering
    \includegraphics[width=.85\linewidth]{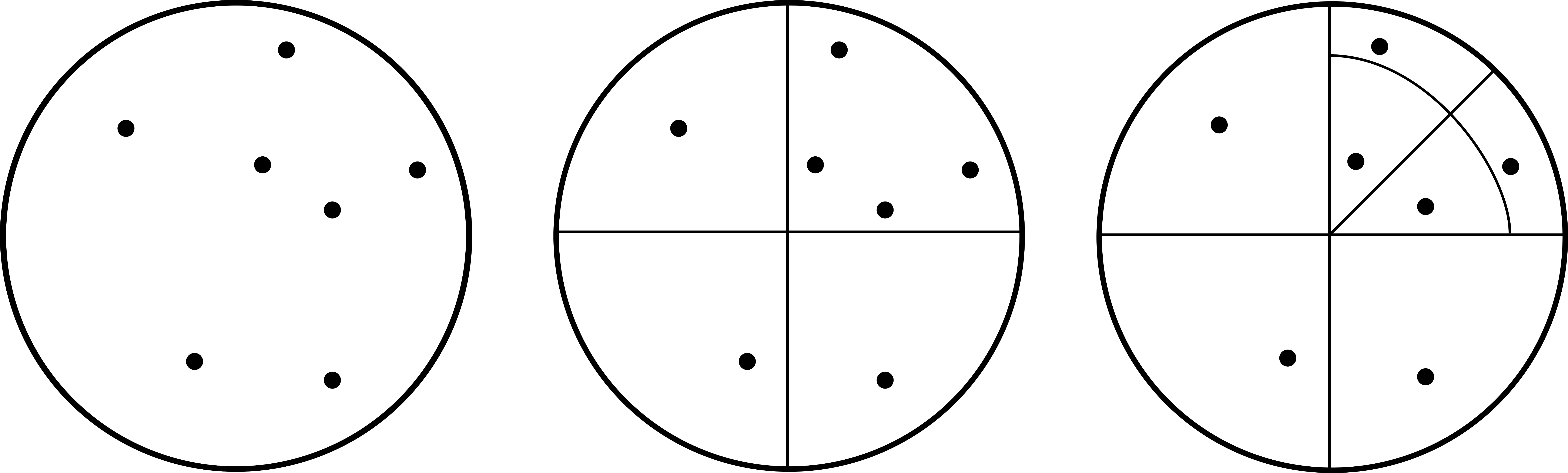}
    \caption{Building a polar quadtree: Poincar\'e disk with seven points, initially split into four pie slices, and splitting one of them along the angular and radial direction.}
    \label{fig:polarQuadtree}
\end{figure}

In hyperbolic space, data structures have to be adjusted to fit the specifics of the space.
A possible translation of the quadtree, used by the Barnes-Hut acceleration of t-SNE in Euclidean space, to hyperbolic space, is the polar quadtree data structure~\cite{looz2015generating}.
The root cell in this case is not a square or rectangle that encompasses all points, as in the Euclidean case, but a circle in the Poincar\'e disk that includes all input points.
This circle is then split along the angular and radial directions, to form polar quadrilaterals as cells, see \cref{fig:polarQuadtree}.

Denoting the angular direction by~$\phi$, a split in this direction is performed at~${\midd_\phi=(\max_\phi+\min_\phi)/2}$, where~$\max_\phi$ and~$\min_\phi$ are the respective maximal and minimal angular values of the current cell.
For each of the resulting four sub-cells to represent the same area in hyperbolic space, a split in radial direction~$r$ is performed at
\begin{align}
\label{equ:EqualAreaSplitting}
    \midd_r = \acosh\left(\frac{\cosh(\max_r)+\cosh(\min_r)}{2}\right).
\end{align}
It can be shown that inserting a node into this tree takes~$\mathcal{O}(\log n)$ time, when~$n$ nodes are present in the tree~\cite[Sec.~3.2]{looz2015generating}.
The polar quadtree data structure was originally introduced for the fast generation of random hyperbolic graphs.
We will use it to translate the Barnes-Hut acceleration of t-SNE to hyperbolic space.

\section{Related Work}
\label{sec:RelatedWork}

Methods for dimensionality reduction can be classified according to whether their embedding is obtained linearly or non-linearly and whether they aim to preserve local or global distances.
Here, we focus on t-SNE~\cite{vanDerMaaten2008visualizing}, which is a non-linear, locally preserving method.
Other methods in this class include LLE (Locally Linear Embedding)~\cite{roweis2000nonlinear}, LE (Laplacian Eigenmaps)~\cite{belkin2001laplacian}, LAMP (Local Affine Multidimensional Projection)~\cite{joia2011local}, and UMAP~\cite{becht2019dimensionality}.
We refer to a recent survey for the advantages and disadvantages of the respective classes and methods~\cite{xia2021revisiting}.
The survey states that non-linear embedding techniques, such as t-SNE, ``preserve local neighborhood[s] in [the] D[imensionality]R[eduction] processes''.
Furthermore, they find that t-SNE ``perform[s] the best in cluster identification and membership identification.''
This motivates our focus on t-SNE.

Considering the t-SNE gradient (\cref{equ:KLGradientSplit}), the two sums can be interpreted as attractive and repulsive forces, respectively, acting on the embedding points~$\mathbf{y}_i$.
This interpretation is in direct correspondence to force-directed graph layouts, see~\cite{bohm2022attraction} for a detailed discussion of the spectrum of attractive and repulsive forces, related methods, and the impact on embeddings.
For a detailed discussion, we will focus on two aspects: embeddings to hyperbolic space and acceleration structures for t-SNE.

\subsection{Hyperbolic Embeddings}
\label{sec:HyperbolicEmbeddings}

Embeddings into hyperbolic space have been studied widely in the context of graph embeddings.
Here, it was shown to be possible to embed trees into two-dimensional hyperbolic space with arbitrarily low distortion~\cite{sarkar2011low}. 
Additionally, many real-world graphs and networks have properties that can be modeled using hyperbolic geometry. 
Random graphs created in the hyperbolic plane~\cite{krioukov2010hyperbolic,friedrich2015cliques} exemplify some of these properties, such as the power-law degree distribution, small diameter, and high clustering, similar to those observed in real-world networks. 
Examples of such graphs are the internet~\cite{boguna2010sustaining} and social network connections~\cite{verbeek2014metric}.
Recent works introduced embedding techniques that scale well to large networks~\cite{nickel2017poincare, blasius2018efficient, sala2018representation} and show the relevance of hyperbolic space for biological data~\cite{zhou2018hyperbolic}.

Subsequently, research started to investigate the potential of hyperbolic spaces for embedding high-dimensional data sets without graph structures.
In this area, several works study extensions of multidimensional scaling (MDS)~\cite{ingram2008glimmer} to hyperbolic space (h-MDS)~\cite{walter2004h,cvetkovski2011multidimensional} as well as extensions of self-organizing maps~\cite{kohonen1982self} to the hyperbolic setting~\cite{ontrup2006large}.
By comparing MDS embeddings of high-dimensional data into Euclidean and hyperbolic space, it was found that the latter resulted in less metric distortion~\cite{sala2018representation}.
This suggests that hierarchical, high-dimensional data, similar to large networks as discussed above, follow an intrinsic hyperbolic metric structure~\cite[Thm.~1]{lin2023hyperbolic}.
On the flip side, many high-dimensional data sets, like networks and graphs, but also single-cell RNA sequencing measurements are of a hierarchical nature~\cite{wu2020accounting}, which spurred the interest for dimensionality reduction~\cite{klimovskaia2020poincare}.

Several extensions of t-SNE to hyperbolic space have been proposed.
The Cauchy Origin-SNE (CO-SNE)~\cite{guo2022co} starts by interpreting the high-dimensional data as hyperbolic by computing the probability distribution~$P$ via the Riemannian normal distribution.
Furthermore, the low-dimensional probabilities~$Q$ are derived using the Cauchy distribution.
Additionally, to preserve hierarchical structures, the cost function (\cref{equ:KLDivergence}) has an additional term to help preserve the distances between the high-dimensional points and the origin.

An alternative extension of t-SNE to hyperbolic space is given by the Poincar\'e maps~\cite{klimovskaia2020poincare}.
Here, the starting point is a nearest-neighbor graph on the high-dimensional data to which additional edges are added until the entire data set is represented by one connected component.
The weights on the edges are modeled by a Gaussian kernel.
The high-dimensional probabilities~$P$ are given by the Relative Forest Accessibility matrix on the graph, while the low-dimensional probabilities~$Q$ are provided by Gaussian kernels.
As a cost function, a symmetric version of the Kullback-Leibler divergence is used.

A third and final extension of t-SNE to hyperbolic space is hyperbolic SNE (h-SNE)~\cite{zhou2021hyperbolic}.
The cost function is enhanced with a term to increase the sensitivity to large distance values, as proposed in g-SNE~\cite{zhou2018using}.
A hyperbolic embedding is obtained by replacing the Euclidean distance terms in the gradient with hyperbolic distances.
As a limitation, the authors identify the lack of any acceleration scheme and thereby the limitation on the size of the data set that can be embedded.
Their embedding scheme performs well until a data set size of about~6,000 points~\cite[p.~11]{zhou2021hyperbolic}.

In conclusion, several extensions of t-SNE to hyperbolic space have been proposed.
These alter the algorithm to accommodate different aspects of the embedding.
Yet, all suffer from the lack of acceleration structures and thus turn to sampling the data before the embedding or performing stochastic approximations of the gradient descent.
In this paper, we present an acceleration data structure suitable to speed up the computation of hyperbolic t-SNE embeddings.

\subsection{Accelerating t-SNE}
\label{sec:Acceleratingt-SNE}

In \cref{sec:BarnesHutAcceleration}, we discussed the Barnes-Hut acceleration method for Euclidean t-SNE.
Several alternatives for this acceleration are available. 
One of these uses the principle of Fourier transforms~\cite{linderman2019fast}.
For this approach, the embedding domain is covered with a regular grid, and the probability distribution~$Q$ is computed at the grid points instead of at the points~$\mathbf{y}_i$.
The second term of \cref{equ:KLGradientSplit} is then interpolated between the grid values.

A similar approach rewrites \cref{equ:KLGradient} in terms of a scalar field representing the point density and a vector field representing the forces, both acting on the regular grid points~\cite{pezzotti2019gpgpu}.
This enables the use of parallelized graphics hardware to solve the embedding problem with linear complexity, assuming that the grid size is~$\ll n,$ with~$n$ points to be embedded.
Furthermore, this can be combined with the quadtree approach, by building a dual-hierarchy setup on both the embedding and its field representation~\cite{vanDeRuit2021efficient}.
This approach provides a complexity of~$\mathcal{O}(n)$ while significantly reducing the number of interactions between the hierarchies, compared to the other accelerations.

One difficulty with embedding into hyperbolic space is that it is not a linear space. 
For instance, it is not possible to zoom in on a part of the hyperbolic space without changing the fundamental structure of the embedding~\cite{eppstein2021limitations}.
Both the Fourier transforms and the vector field approach need a regular grid representation of the embedding space, but without a uniform scaling, there is only one fixed resolution of such available in hyperbolic space.
Similarly, it is not possible to translate hierarchies with congruent cells on a level and similar cells across levels, like Barnes-Hut~\cite{vanDerMaaten2014accelerating} and the dual quadtrees~\cite{vanDeRuit2021efficient}, as these require a similar, uniform rescaling.
Therefore, efficient computations of embeddings in hyperbolic space must use the geometric structures of hyperbolic space to their advantage~\cite{keller2020hydra}.
We aim to address this with our hierarchical acceleration data structure that we present in the following.


\section{A Hierarchical Acceleration Structure for Hyperbolic t-SNE}
\label{sec:AHierarchicalAccelerationStructureForHyperbolicT-SNE}

Our proposed solution for accelerating hyperbolic t-SNE embeddings is based on a data structure, which we describe in \cref{sec:AModifiedPolarQuadtreeForEmbeddingAcceleration}, that is adjusted to hyperbolic space.
We then proceed to explain how the data structure can be used to approximate the hyperbolic gradient and thereby speed up the gradient descent steps of the optimization (\cref{sec:ApproximatingTheHyperbolicGradient}).
Finally, before going to some experimental validations, we will investigate how our approach can be used to accelerate the different variants of hyperbolic t-SNE (\cref{sec:RelationsOfTheAccelerationToPreviousApproaches}).

\subsection{Modified Polar Quadtree -- Embedding Acceleration}
\label{sec:AModifiedPolarQuadtreeForEmbeddingAcceleration}

\begin{figure}
    \centering
    \includegraphics[width=0.65\linewidth]{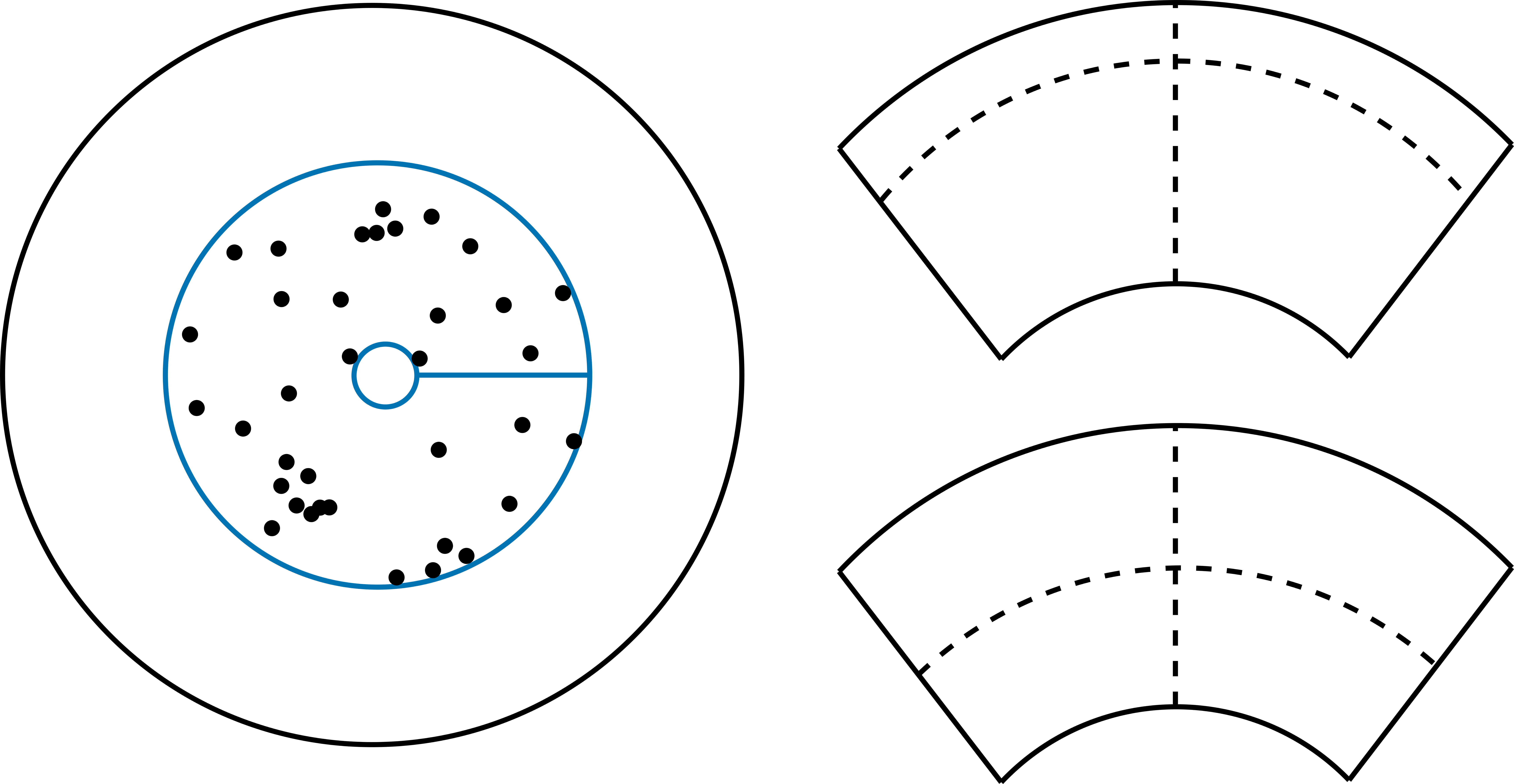}
    \hfill
    \def\svgwidth{.32\linewidth}
\begingroup%
  \makeatletter%
  \providecommand\color[2][]{%
    \errmessage{(Inkscape) Color is used for the text in Inkscape, but the package 'color.sty' is not loaded}%
    \renewcommand\color[2][]{}%
  }%
  \providecommand\transparent[1]{%
    \errmessage{(Inkscape) Transparency is used (non-zero) for the text in Inkscape, but the package 'transparent.sty' is not loaded}%
    \renewcommand\transparent[1]{}%
  }%
  \providecommand\rotatebox[2]{#2}%
  \newcommand*\fsize{\dimexpr\f@size pt\relax}%
  \newcommand*\lineheight[1]{\fontsize{\fsize}{#1\fsize}\selectfont}%
  \ifx\svgwidth\undefined%
    \setlength{\unitlength}{3.68467091bp}%
    \ifx\svgscale\undefined%
      \relax%
    \else%
      \setlength{\unitlength}{\unitlength * \real{\svgscale}}%
    \fi%
  \else%
    \setlength{\unitlength}{\svgwidth}%
  \fi%
  \global\let\svgwidth\undefined%
  \global\let\svgscale\undefined%
  \makeatother%
  \begin{picture}(1,1)%
    \lineheight{1}%
    \setlength\tabcolsep{0pt}%
    \put(0,0){\includegraphics[width=\unitlength,page=1]{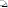}}%
    \put(0.10953678,0.73073101){\color[rgb]{0,0,0}\makebox(0,0)[lt]{\lineheight{1.25}\smash{\begin{tabular}[t]{l}$\mathbb{D}$\end{tabular}}}}%
  \end{picture}%
\endgroup%

    \caption{
        Left: Initial annulus root cell of a polar quadtree.
        Center: Splitting the polar quadrilaterals such that they represent equal hyperbolic area (top) or at the center of the embedding coordinates (bottom). 
        Right: Two polar quads that have different elements as their respective longest internal distances, highlighted in blue.
    }
    \label{fig:polarQuadtreeAnulusSplittingAndExtremes}
\end{figure}

We aim at building a hierarchical data structure corresponding to the Barnes-Hut tree for t-SNE~\cite{vanDerMaaten2014accelerating} but designed for the Poincar\'e disk model of hyperbolic space.
Recall that the Barnes-Hut tree starts from an initial quadrilateral encompassing all points and is built hierarchically by splitting the quadrilateral into four congruent quadrilaterals.
Because of the curved nature of hyperbolic space, hierarchical splitting into congruent tiles is not possible.
There exist tilings of hyperbolic space with congruent tiles~\cite{conway2008symmetries}, see an example in \cref{fig:PoincareDisk}, but these do not support a hierarchy built from similar tiles.
Therefore, we abandon congruent tiles and instead settle for a hierarchy that consists of similar tiles both laterally at one level of the hierarchy and across different levels.
We achieve this by starting from an annulus on the Poincar\'e disk, see \cref{fig:polarQuadtreeAnulusSplittingAndExtremes} left, as the root node of our version of a polar quadtree.
This annulus contains all embedding points~$\mathbf{y}_i$ and is split into four similar polar quadrilaterals by cutting in the radial and angular directions.

In the Euclidean case, when cutting a quadrilateral into four congruent smaller quadrilaterals, each of these has the same diagonal length.
Thus, the maximum distance of two points in one cell of a quadtree shrinks uniformly by a factor of~$0.5$ from one level of the hierarchy to the next. 
There are two important differences when going to the hyperbolic setting. 
First, the longest distance within a polar quadrilateral is not necessarily the diagonal.
While this is true for polar quadrilaterals close to the origin, polar quadrilaterals towards the outside of the disk have the longest distance along their outer arc, see \cref{fig:polarQuadtreeAnulusSplittingAndExtremes} right.
Hence, when checking for the largest possible difference between points within one cell of the polar quadtree, we have to check not only the diagonal but also one of the radial and one of the polar edges of the quadrilateral.
Second, as the polar quadrilaterals on one level of the hierarchy are not congruent anymore, they can exhibit different longest distances within them.
Thus, there is no unique shrinkage factor across the levels of the hierarchy. 
However, preliminary experiments suggest that the shrinkage factor approaches~$0.5$ rapidly, after just a few levels of the hierarchy.

\begin{figure}
    \centering
    \includegraphics[width=1.\linewidth]{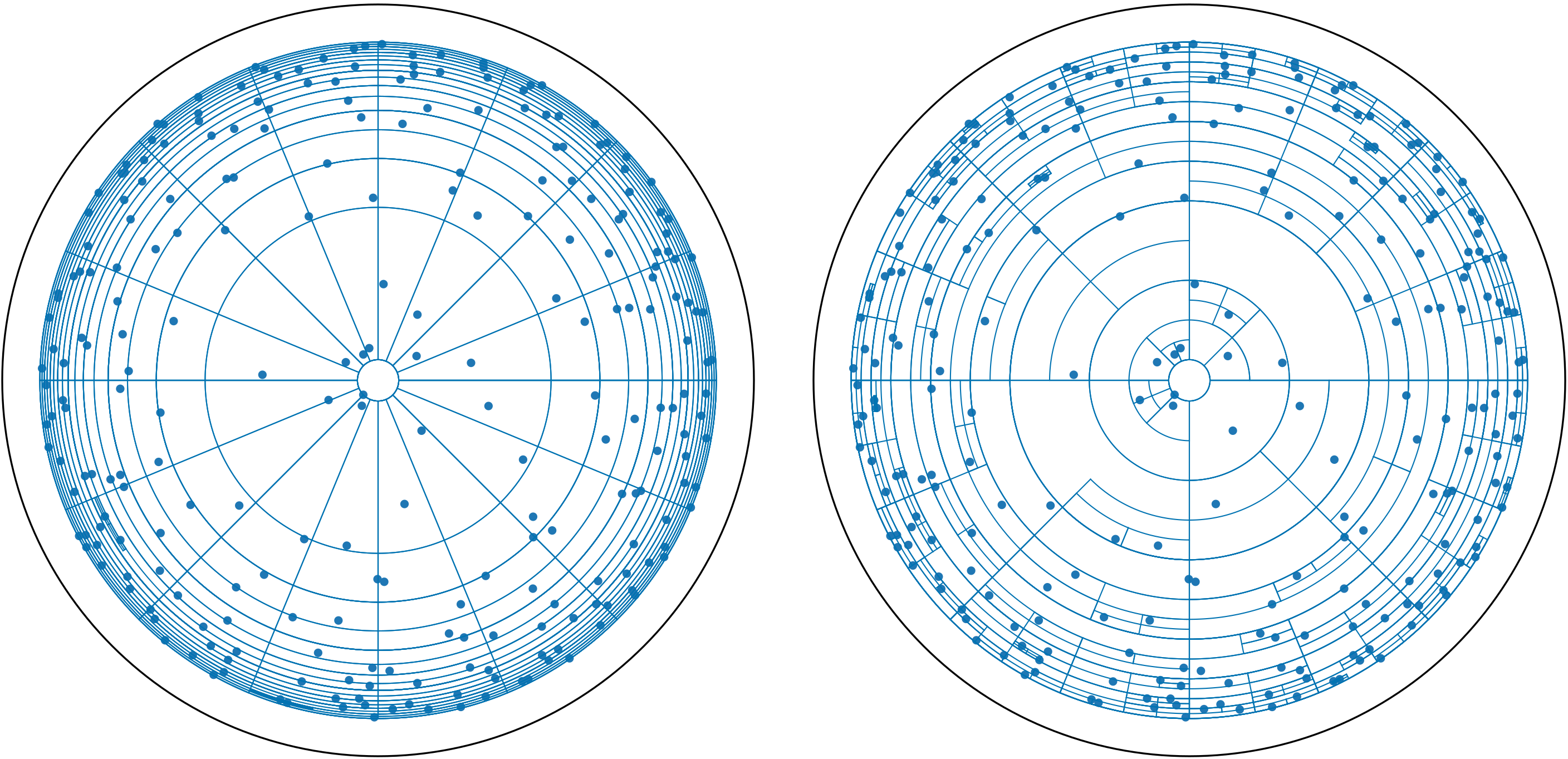}
    \caption{
        Effect of the two splitting choices on the polar quadtree, note the long pieces towards the center when splitting according to \cref{equ:EqualAreaSplitting} on the left, while cells are more compact when splitting according to \cref{equ:EqualEmbeddingLengthsSplitting} on the right.
    }
    \label{fig:polarQuadtreeSplitting}
\end{figure}

As stated above, we choose an annulus including all embedded points~$\mathbf{y}_i$ as root node, see \cref{fig:polarQuadtreeAnulusSplittingAndExtremes} left.
So far, this approach has been following the polar quadtree construction as outlined in \cref{sec:PolarQuadtree}.
However, when splitting cells in the polar quadtree, we strive for a tree that best supports the subsequent approximation scheme, therefore we divert from the original splitting procedure~\cite{looz2015generating} and alter it to better adapt to our case of data embedding.
We continue splitting in the middle of the angular direction, that is at~${\midd_\phi=(\max_\phi+\min_\phi)/2}$.
However, in our experiments, we observed that splitting along the radial direction according to \cref{equ:EqualAreaSplitting} creates larger quadrilateral cells towards the center of the Poincar\'e disk, see \cref{fig:polarQuadtreeSplitting}.
As t-SNE begins with a PCA initialization placing all~$\mathbf{y}_i$ close to the disk center~\cite{kobak2021initialization}, this way of building the polar quadtree does not provide a good resolution, especially for these first iterations.
Therefore, we propose a different splitting rule that creates more equal-sized quadrilaterals from the perspective of looking at the Poincar\'e disk (\cref{fig:polarQuadtreeAnulusSplittingAndExtremes} center):
\begin{align}
\label{equ:EqualEmbeddingLengthsSplitting}
    \midd_r = \frac{\max_r+\min_r}{2}.
\end{align}

We will evaluate the effect of this new splitting choice on the performance of the approximation in \cref{sec:TimeGainBySplittingChoices}.
Note that following this splitting choice leads to differently shaped nodes and thus a different tree than the original polar quadtree~\cite{looz2015generating}.

When building the polar quadtree, in addition to the coordinates of each polar quadrilateral at the nodes, we also store one piece of additional information.
In the leaf nodes, we store the single point~$\mathbf{y}_i$ located within the polar quadrilateral of this node.
In all other nodes, we store a midpoint~$\mathbf{y}_{cell}$ of all points~$\mathbf{y}_i$ in this cell.
In the Euclidean setting, this is simply the arithmetic mean of the stored embedding points, which is not available in hyperbolic space. 
There, the midpoint is given by the Fr\'echet mean, which is defined indirectly as the solution to a variance-minimization problem.
This could be solved to an~$(1-\varepsilon)$-approximation by iteratively solving an optimization problem for every cell during tree construction~\cite{cao2023poincare}.
To avoid this, we turn to a pseudo-Fr\'echet mean that has the closed form
\begin{align}
\label{equ:EinsteinMidpoint}
    m\left(\{\mathbf{v}_j\}\right) = \sum_{j}\left(\frac{\gamma(\mathbf{v}_j)}{\sum_{\ell}\gamma(\mathbf{v}_\ell)}\right)\mathbf{v}_j,
\end{align}
where $\gamma(\mathbf{v}_j)=1/\sqrt{1-\left\|\mathbf{v}_j\right\|^2}$ and ~$\mathbf{v}_j$ are the coordinates of~$\mathbf{y}_j$ interpreted in the Klein model of hyperbolic space, which can easily be computed~\cite{gulcehre2018hyperbolic}.
This is only an approximation of the midpoint and comes with an error rate of about~7\% with regard to the Fr\'echet variance problem~\cite[Appendix~H]{lou2020differentiating}, however, it enables us to compute the midpoint as a rolling average.
That is, we can build the tree by successively adding points and updating the cell midpoints on the fly, which means that inserting a new point and updating all midpoint information still has~$\mathcal{O}(\log(n))$ cost.

\subsection{Approximating the Hyperbolic Gradient}
\label{sec:ApproximatingTheHyperbolicGradient}

There are different possible ways of adapting the objective of the Euclidean t-SNE for hyperbolic embeddings. 
For our experiments, we use an objective that resembles the Euclidean case as closely as possible.
Therefore, we keep the high-dimensional probabilities (\cref{equ:HighDimensionalProbability}).
Replacing the Euclidean distance in the low-dimensional probabilities (\cref{equ:LowDimensionalProbability}) gives
\begin{align}
\label{equ:LowDimensionalPorbabilityHyperbolic}
    q^\mathcal{H}_{ij} = \frac{\left(1+(d^\mathcal{H}_{ij})^2\right)^{-1}}{\sum_{k\neq\ell}\left(1+(d^\mathcal{H}_{ij})^2\right)^{-1}},
\end{align}
with~$d^\mathcal{H}_{ij}$ the hyperbolic distances (\cref{equ:PoincareDistance}) of~$\mathbf{y}_i$ and~$\mathbf{y}_j$.
Thereby, in the gradient of the cost function, we need the variation of the hyperbolic distance
\begin{align}
\label{equ:PoinicareDistanceDerivative}
    \frac{\delta d^\mathcal{H}_{ij}}{\delta\mathbf{y}_i} = \frac{4 ((\left\|\mathbf{y}_j\right\|^2-2\langle\mathbf{y}_i,\mathbf{y}_j\rangle+1)\mathbf{y}_i/\alpha - \mathbf{y}_j)\  
    }{\alpha\beta\sqrt{\gamma^2-1}},
\end{align}
where $\langle.,.\rangle$ denotes the standard inner product and~$\left\|.\right\|$ the standard norm of~$\mathbb{R}^2$.
In addition, ${\alpha=1-\left\|\mathbf{y}_i\right\|^2}$, ${\beta=1-\left\|\mathbf{y}_j\right\|^2}$, and ${\gamma=1+\frac{2}{\alpha\beta}\left\|\mathbf{y}_i-\mathbf{y}_j\right\|^2}$.
The hyperbolic gradient is the product of~$\lambda_{\mathbf{y}_i}^{-1}$ from \cref{equ:PoincareMetric} with the variation
\begin{align}
\label{equ:KLGradientHyperbolic}
    \frac{\delta C^\mathcal{H}}{\delta\mathbf{y}_i} = 4\sum_{j\neq i} (p_{ij}-q^\mathcal{H}_{ij})(1+{d^\mathcal{H}_{ij}}^2)^{-1}\frac{\delta {d^\mathcal{H}_{ij}}}{\delta\mathbf{y}_i}.
\end{align} 
The derivation of the gradient is analog to that of Co-SNE~\cite{guo2022co} and h-SNE~\cite{zhou2018hyperbolic}.
Similar to the Euclidean case, we can rewrite the variations in a split form of two sums as
\begin{align}
\label{equ:KlGradientHyperbolicSplit}
    4 \left(\sum_{j\neq i} p_{ij}q^\mathcal{H}_{ij}Z^\mathcal{H}\frac{\delta {d^\mathcal{H}_{ij}}}{\delta\mathbf{y}_i}-\sum_{j\neq i}\left(q^\mathcal{H}_{ij}\right)^2 Z^\mathcal{H}\frac{\delta {d^\mathcal{H}_{ij}}}{\delta\mathbf{y}_i}\right),
\end{align}
where~$Z^\mathcal{H}=\sum_{k\neq\ell}\left(1+{d^\mathcal{H}_{ij}}^2\right)^{-1}$.

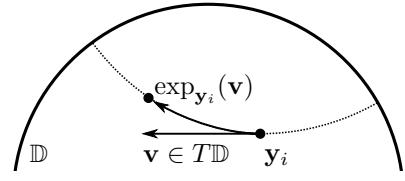
\begin{figure}
    \centering
    \def\svgwidth{.6\linewidth}
\begingroup%
  \makeatletter%
  \providecommand\color[2][]{%
    \errmessage{(Inkscape) Color is used for the text in Inkscape, but the package 'color.sty' is not loaded}%
    \renewcommand\color[2][]{}%
  }%
  \providecommand\transparent[1]{%
    \errmessage{(Inkscape) Transparency is used (non-zero) for the text in Inkscape, but the package 'transparent.sty' is not loaded}%
    \renewcommand\transparent[1]{}%
  }%
  \providecommand\rotatebox[2]{#2}%
  \newcommand*\fsize{\dimexpr\f@size pt\relax}%
  \newcommand*\lineheight[1]{\fontsize{\fsize}{#1\fsize}\selectfont}%
  \ifx\svgwidth\undefined%
    \setlength{\unitlength}{388.51501465bp}%
    \ifx\svgscale\undefined%
      \relax%
    \else%
      \setlength{\unitlength}{\unitlength * \real{\svgscale}}%
    \fi%
  \else%
    \setlength{\unitlength}{\svgwidth}%
  \fi%
  \global\let\svgwidth\undefined%
  \global\let\svgscale\undefined%
  \makeatother%
  \begin{picture}(1,0.42859345)%
    \lineheight{1}%
    \setlength\tabcolsep{0pt}%
    \put(0.33808899,0.02094389){\color[rgb]{0,0,0}\makebox(0,0)[lt]{\lineheight{1.25}\smash{\begin{tabular}[t]{l}$\mathbf{v}\in T\mathbb{D}$\end{tabular}}}}%
    \put(0,0){\includegraphics[width=\unitlength,page=1]{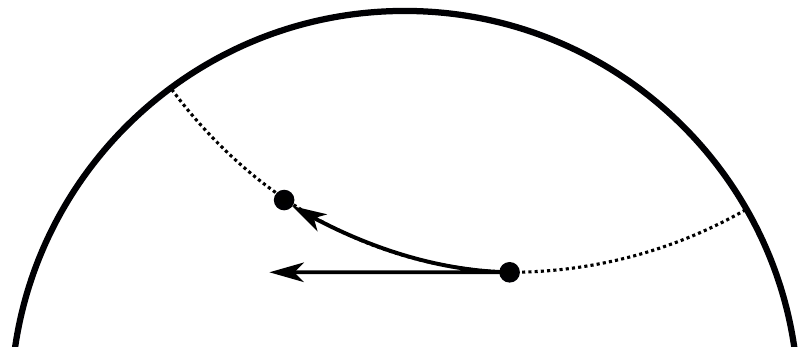}}%
    \put(0.0553173,0.02092155){\color[rgb]{0,0,0}\makebox(0,0)[lt]{\lineheight{1.25}\smash{\begin{tabular}[t]{l}$\mathbb{D}$\end{tabular}}}}%
    \put(0.36580433,0.19067119){\color[rgb]{0,0,0}\makebox(0,0)[lt]{\lineheight{1.25}\smash{\begin{tabular}[t]{l}$\exp_{\mathbf{y}_i}(\mathbf{v})$\end{tabular}}}}%
    \put(0.64006216,0.02527608){\color[rgb]{0,0,0}\makebox(0,0)[lt]{\lineheight{1.25}\smash{\begin{tabular}[t]{l}$\mathbf{y}_i$\end{tabular}}}}%
  \end{picture}%
\endgroup%

    \caption{
        The exponential map projects a step~$\mathbf{v}$ at point~$\mathbf{y}_i$ in the tangent space~$T_{\mathbf{y}_i}\mathbb{D}$ onto~$\mathbb{D}$.
        Thus, a step follows a straight line in the Poincar\'e disk, as shown in \cref{fig:PoincareDisk}.
    }
    \label{fig:ExponentialMap}
\end{figure}

To ensure each gradient descent-step taken from~$\mathbf{y}_i$ in the Euclidean tangent space~$T_{\mathbf{y}_i}\mathbb{D}$ of~$\mathbb{D}$ is projected to the correct point on the Poincar\'e disk, a standard procedure is to utilize the exponential map, see \cref{fig:ExponentialMap}.
That is, for a Euclidean direction~${\mathbf{v}\in\mathbb{R}^2}$, we project the corresponding step taken from a point~$\mathbf{y}_i$ in the Poincar\'e disk to
\begin{align}
    \exp_{\mathbf{y}_i}(\mathbf{v}) = \mathbf{y}_i \oplus \left(\tanh\left(\frac{\lambda_{\mathbf{y}_i}\left\|\mathbf{v}\right\|}{2}\right)\frac{\mathbf{v}}{\left\|\mathbf{v}\right\|}\right),
\end{align}
with~$\lambda_{\mathbf{y}_i}$ from \cref{equ:PoincareMetric} and~$\oplus$ the M\"obius addition, which for two points~${\mathbf{y}_i,\mathbf{y}_j\in\mathbb{D}}$ is defined as:
\begin{align}
    \mathbf{y}_i \oplus \mathbf{y}_j = \frac{(1+2\langle\mathbf{y}_i,\mathbf{y}_j\rangle + \left\|\mathbf{y}_j\right\|^2)\mathbf{y}_i + (1-\left\|\mathbf{y}_i\right\|^2)\mathbf{y}_j}{1+2\langle\mathbf{y}_i,\mathbf{y}_j\rangle+\left\|\mathbf{y}_i\right\|^ 2\left\|\mathbf{y}_j\right|^2}.
\end{align}
See~\cite{ganea2018hyperbolic} for a more general version incorporating varying curvature of the hyperbolic space.

Performing gradient descent close to the edge of the Poincar\'e disk can move points outside of the disk.
We follow the previously suggested solution of projecting the points back to the strict interior of the Poincar\'e disk after each gradient step~\cite[p.~5]{nickel2017poincare}:
\begin{align}
\label{equ:HyperbolicProjection}
    \proj(\mathbf{y}_i) = 
    \begin{cases} 
        \mathbf{y}_i/\left\|\mathbf{y}_i\right\|-\varepsilon & \text{ if }\left\|\mathbf{y}_i\right\|\geq1\\ 
        \mathbf{y}_i & \text{otherwise}
    \end{cases}.
\end{align}

The polar quadtree with our modified splitting rule \cref{equ:EqualEmbeddingLengthsSplitting} will serve as the main acceleration tool to speed up the evaluation of the hyperbolic gradient as given in \cref{equ:KlGradientHyperbolicSplit}.
Note that, just as in the Euclidean case, for a sparse high-dimensional probability distribution~$P$ with truncated Gaussians in \cref{equ:HighDimensionalProbability}, the first sum of \cref{equ:KlGradientHyperbolicSplit} can be evaluated without negatively affecting the algorithm performance.
To speed up the computation of the second sum of \cref{equ:KlGradientHyperbolicSplit}, we proceed analogously to the Barnes-Hut approach for Euclidean t-SNE~\cite{vanDerMaaten2014accelerating}.
That is, we observe that if a cell of the polar quadtree is sufficiently small and sufficiently far away from a point~$\mathbf{y}_i$, the contributions~${-\left(q^\mathcal{H}_{ij}\right)^2Z^\mathcal{H}{\delta d^\mathcal{H}_{ij}/\delta\mathbf{y}_i}}$ will be similar for all points~$\mathbf{y}_j$ inside this cell.
Therefore, we replace these summands by
\begin{align}
\label{equ:CellApproximation}
    -N_{cell} \left(q^\mathcal{H}_{i,cell}\right)^2 Z^\mathcal{H} \frac{\delta d^{\mathcal{H}}(\mathbf{y}_i,\mathbf{y_{cell})}}{\delta\mathbf{y}_i},
\end{align}
where~$N_{cell}$ is the number of points~$\mathbf{y}_j$ in the cell, $\mathbf{y}_{cell}$ is the midpoint of the cell according to \cref{equ:EinsteinMidpoint}, and 
\begin{align*}
    q^\mathcal{H}_{i,cell} Z^\mathcal{H} = \left(1 + d^\mathcal{H}(\mathbf{y}_i,\mathbf{y}_{cell})^2\right)^{-1}.
\end{align*}
When evaluating the second sum in \cref{equ:KLGradientSplit} for a point~$\mathbf{y}_i$, we perform a depth-first traversal of the polar quadtree.
At each node, we check the condition~$r_{cell}/d^\mathcal{H}(\mathbf{y}_i,\mathbf{y}_{cell})<\theta$, the hyperbolic analog of \cref{equ:CellApproximationCriterion}, and if it holds, we cull the subtree and replace its summands by an approximation according to \cref{equ:CellApproximation}.
See \cref{fig:teaser}, left, for an illustration of the approximation, similar to the Euclidean illustration in \cref{fig:BarnesHutApproximation}.
We will evaluate the effectiveness of this approximation and its effects on the embedding quality in \cref{sec:Experiments}.

\subsection{Application to other hyperbolic t-SNE Schemes}
\label{sec:RelationsOfTheAccelerationToPreviousApproaches}

This approach of translating the Barnes-Hut approximation data structure to hyperbolic space enables the acceleration of t-SNE embeddings in hyperbolic spaces.
Note that the data structure and the approach described here are not contradictory but rather complementary to previous hyperbolic variants of t-SNE~\cite{klimovskaia2020poincare,zhou2021hyperbolic,guo2022co}.
All these methods can be augmented by our data structure to efficiently compute the t-SNE gradient (\cref{equ:KLGradientSplit}) and thus provide faster results.
In that sense, we provide a new building block for hyperbolic dimensionality reduction.
Here, we briefly discuss the gradients of the methods~\cite{zhou2021hyperbolic,guo2022co} to discuss how our acceleration can be implemented there.

Hyperbolic SNE~\cite{zhou2021hyperbolic} uses the cost function
\begin{align*}
    C+\lambda\hat{C}=\KL(P||Q)+\lambda \KL(\hat{P}||\hat{Q})
\end{align*}
 from~\cite{zhou2018using}, where~${\lambda\in\mathbb{R}}$  is a weighting parameter and
\begin{align*}
    \hat{p}_{ij} = \frac{1+\left\|\mathbf{x}_i-\mathbf{x}_j\right\|^2}{\sum_{k\neq\ell}(1+\left\|\mathbf{x}_k-\mathbf{x}_\ell\right\|^2)},&&
    \hat{q}_{ij} = \frac{1+\left\|\mathbf{y}_i-\mathbf{y}_j\right\|^2}{\sum_{k\neq\ell}(1+\left\|\mathbf{y}_k-\mathbf{y}_\ell\right\|^2)}.
\end{align*}
In the hyperbolic case, the variations~${\delta(C+\lambda\hat{C})/\delta\mathbf{y}_i}$ are 
$4\sum_{j\neq i}(p_{ij}-q_{ij})\left(1+\left\|\mathbf{y}_i-\mathbf{y}_j\right\|^2\right)^{-1}(\mathbf{y}_i-\mathbf{y}_j)
    -4\lambda\sum_{j\neq i} (p_{ij}-q^\mathcal{H}_{ij})\left(1+(d^\mathcal{H}_{ij})^2\right)^{-1}\frac{\delta d^\mathcal{H}_{ij}}{\delta\mathbf{y}_i},
$
where the first part corresponds to a scaled version of \cref{equ:KLGradient} and the second part corresponds to \cref{equ:KLGradientHyperbolic}.
Thus, both can be accelerated, respectively.

CO-SNE\cite{guo2022co} uses the cost function
$
    \lambda_1 C + \lambda_2 H = \lambda_1 \KL(P||Q) + \frac{\lambda_2}{n}\sum_{i=1}^n \left(\left\|\mathbf{x}_i\right\|^2 - \left\|\mathbf{y}_i\right\|^2\right)^2
$
with the variations~${\delta(\lambda_1 C + \lambda_2 H)/\delta\mathbf{y}_i}$ equal to
$
    4\lambda_1\sum_{j\neq i} (p_{ij}-q^\mathcal{H}_{ij})\left(1+(d^\mathcal{H}_{ij})^2\right)^{-1}\frac{\delta d^\mathcal{H}_{ij}}{\delta\mathbf{y}_i}
    +\frac{4\lambda_2}{n}\left(\left\|\mathbf{x}_i\right\|^2 - \left\|\mathbf{y}_i\right\|^2\right)\mathbf{y}_i,
$
where the first part can be rewritten equal to \cref{equ:KLGradientHyperbolic,equ:KlGradientHyperbolicSplit} while the second part does not need any acceleration as it can be evaluated in constant time.

Note that the gradient used by the Poincar\'e maps approach~\cite{klimovskaia2020poincare} is not explicitly given in the publication.
The derivation of the gradient is outside the scope of this publication. 
Still, at least the first summand can be rewritten equivalently to \cref{equ:KlGradientHyperbolicSplit} and the second symmetric summand, can either be rewritten similarly or approximated otherwise.
This shows that our method is versatile in the sense that it provides a building block to integrate into existing hyperbolic t-SNE implementations.


\begin{figure*}
    \includegraphics[width=1.\textwidth]{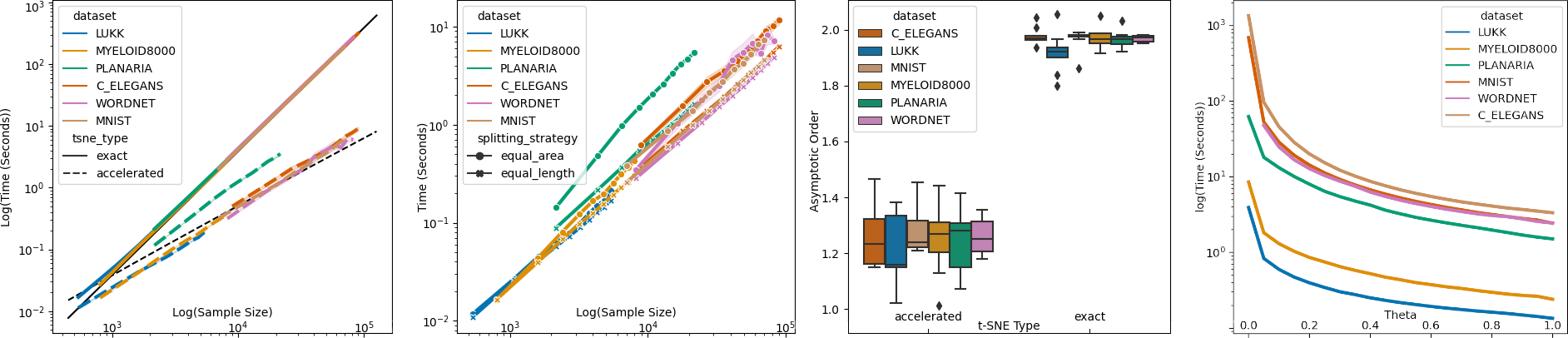}
    \caption{
        Evaluations from left to right:
        (1) 
            Run time behavior of the exact hyperbolic t-SNE embedding vs.\@ the accelerated version on various input data sizes.
            For each data set, measurements are connected as a line; black trend lines for~$\mathcal{O}(n^2)$ and~$\mathcal{O}(n\log_2(n))$ are fitted to the data by regression.
        (2)
            Run time behavior of equal area splitting (\cref{equ:EqualAreaSplitting}) vs lengths splitting (\cref{equ:EqualEmbeddingLengthsSplitting}).
        (3)
            Estimating asymptotic run time of the accelerated and the exact hyperbolic t-SNE embeddings.
        (4)
            Run time behavior of accelerated hyperbolic t-SNE embeddings for different values of~$\theta$.
    }
    \label{fig:Comparisons}
\end{figure*}

\section{Evaluation}
\label{sec:Experiments}

In this section, we will experimentally evaluate our hyperbolic acceleration scheme.
For our experiments, we use the data sets listed in \cref{tab:Datasets}.

The first three data sets and the last one contain data obtained from single-cell RNA sequencing~\cite{andrews2021tutorial}.
The data by Lukk et al., the Planaria data set~\cite{plass2018cell}, and the C.Elegans are experimentally obtained gene expression atlases. 
The first contains human cell data, the latter two contain flatworm data.
In contrast, the MyeloidProgenitors data consists of synthetic data, obtained via a boolean gene regulatory network~\cite{krumsiek2011hierarchical}.
These data sets are chosen due to their size variation and their use in previous hyperbolic t-SNE approaches~\cite{zhou2018hyperbolic,klimovskaia2020poincare,guo2022co}.
We further include the MNIST data set, as it is a frequently used test set for dimensionality reduction algorithms.
It contains 70,000 hand-written images of the ciphers 0 to 9.
For the WordNet data set~\cite{fellbaum1998wordnet}, consisting of word relations, we follow the general approach as outlined in previous work~\cite{nickel2017poincare}. 
That is, we trained a network for 400~epochs, after which no significant change occurred.
From this training, we pick the checkpoint with the lowest cost function value as input for our embeddings.

\begin{table}[htb]
    \centering
    \caption{
        Data sets used in the experiments with the number of points, the dimension, and the number of labeled classes.
    }
    \renewcommand*{\arraystretch}{1.2}
    \begin{tabular}{l|l|r|r|r}
        Name                & Data Type   & \# Points   & \# Dim. & \# Cl.\\
        \hline
        \scriptsize LUKK         & single-cell & 5,372       & 369 & 4 \\
        \scriptsize MYELOID8000  & single-cell & 8,000       & 11 & 5 \\
        \scriptsize PLANARIA            & single-cell & 21,612      & 50 & 51 \\
        \scriptsize MNIST               & images      & 70,000      & 784 & 10 \\
        \scriptsize WORDNET             & lexical     & 82,115      & 11 & n/a \\
        \scriptsize C\_ELEGANS           & single cell & 89,701      & 20,222 & 37
    \end{tabular}
    \label{tab:Datasets}
\end{table}

When computing a hyperbolic t-SNE embedding of these data with our method, we perform the following steps that reflect best practices for Euclidean t-SNE as closely as possible.
First, we employ principal component analysis (PCA) to reduce the data to~50 dimensions to speed up computations as previously recommended~\cite[Sec.~4.2]{vanDerMaaten2008visualizing}.
Then, we further employ regular t-SNE strategies by first performing early exaggeration, that is, a series of gradient descent steps for which the attractive forces~$p_{ij}$ are amplified by multiplying them with a factor, see~\cite[Sec.~4.3]{vanDerMaaten2008visualizing}.
We use an exaggeration factor of~12, following a corresponding ablation study~\cite{belkina2019automated}.
We then apply several non-exaggerated gradient descent steps, that is, steps with regular $p_{ij}$ as defined above, as with usual t-SNE~\cite[Sec.~3.4]{vanDerMaaten2008visualizing}.

\begin{table}
    \centering
    \begin{tabular}{l|c|c}
                 & {Accelerated [s]} & {Exact [s]} \\
        \hspace{-0.1cm}Data set & $\min$ / $\avg$ / $\std$ / $\max$ & $\min$ / $\avg$ / $\std$ / $\max$\\
        \hline
        \hspace{-0.1cm}\scriptsize LUKK & 0.13 / 0.17 / 0.03 / 0.46
        & 1.07 / 1.20 / 0.09 / 2.13 \\
        \hspace{-0.1cm}\scriptsize MYELOID8000 & 0.04 / 0.30 / 0.07 / 0.45
        & 1.87 / 2.57 / 0.12 / 2.84 \\
        \hspace{-0.1cm}\scriptsize PLANARIA & 0.73 / 1.62 / 0.29 / 3.35
        & 16.7 / 19.3 / 1.65 / 22.3 \\
        \hspace{-0.1cm}\scriptsize MNIST & 2.33 / 4.57 / 0.27 / 5.48
        & 176. / 191. / 15.4 / 227. \\
        \hspace{-0.1cm}\scriptsize WORDNET & 1.91 / 4.84 / 0.45 / 6.25
        & 245. / 273. / 22.0 / 318. \\
        \hspace{-0.1cm}\scriptsize C\_ELEGANS & 5.28 / 6.31 / 0.42 / 7.47
        & 263. / 316. / 25.0 / 374. \\
    \end{tabular}
    \caption{Run time statistics over five runs for six datasets.}
    \label{tab:RunTimeStatistics}
\end{table}

For the learning rate, we align with a heuristic from the Euclidean case.
In the Euclidean case, the initial learning rate is set to~${\eta=n/12}$~\cite{belkina2019automated}.
We observe that this setting alone causes embeddings that tend to the boundary of the Poincar\'e disk very quickly.
To slow this progression down and to enable a more thorough development of clusters in the hyperbolic case, we modify the Euclidean heuristic and set the initial learning rate to
\begin{align}
    \eta = \frac{n}{12 \cdot 1000}.
\end{align}
This is also because, within the hyperbolic disk, the distance between the left boundary and the right one is not large, for instance, $d^\mathcal{H}\left((-1+10^{-4},0), (1-10^{-4},0)\right)\approx 80$.
Thus, the embedding has to grow significantly slower than in the Euclidean case, where embeddings easily grow to diagonal sizes of several hundred units.
    
From this initial learning rate, we use momentum and gains as described for the Euclidean setting~\cite{vanDerMaaten2008visualizing,jacobs1988increased}.
Gradient descent optimization with momentum is available for hyperbolic space~\cite[Alg.~2]{cho2017riemannian} and we implement it via the machinery discussed above~\cite{ganea2018hyperbolic}.
This allows us to rely on the comparatively rather small initial learning rate that builds up with momentum and gains.
We default to the same parameters used in the Euclidean case, that is, a momentum of~$0.5$ during early exaggeration and a momentum of~$0.8$ during the non-exaggerated iterations.
Furthermore, we run all experiments with a uniform perplexity of~$30$, which is in the recommended range~\cite[Tab.~2]{belkina2019automated} and corresponds to the default values used in previous experiments~\cite{kobak2019art}.
Furthermore, all accelerations use~${\theta=0.5}$, if not specified otherwise, as is used in previous work~\cite{vanDerMaaten2014accelerating}.


\subsection{Time Gain by Acceleration}

As a first set of experiments, we will investigate the time gain obtained by our acceleration.
Here, we are interested in the absolute time gained for each iteration (\cref{sec:ReductionOfTheAbsoluteRunTime}), an estimate of the asymptotic time gain (\cref{sec:EstimatedAsymptoticRunTime}), and the effect of parameter~$\theta$ (\cref{sec:EffectOfThetaOnTheRunTime}) and structural choices (\cref{sec:TimeGainBySplittingChoices}) on the data structure.

\subsubsection{Reduction of the Absolute Run Time}
\label{sec:ReductionOfTheAbsoluteRunTime}

First, we want to measure the effectiveness of our acceleration structure with the following experiment.
For each data set listed in \cref{tab:Datasets}, we consider the ten sample sizes~${n/10,2n/10,\ldots,n}$, where~$n$ is the number of points in the data set.
Then, for each size, we draw five random samples and perform~$250$ iterations of early exaggeration followed by several non-exaggerated iterations of gradient descent.
When running these non-exaggerated iterations, we check regularly whether any point has a distance smaller than~$10^{-4}$ from the boundary of the Poincar\'e disk, measured in embedding coordinates.
If so, we stop the optimization as the embedding has now sufficiently spread across the disk.
At the latest, we always stop the gradient descent after 750 iterations, as in previous studies~\cite{belkina2019automated}

After the optimization, we average the time it took for each iteration across all five random runs.
This provides us with ten different average times per data set, dependent on the size of the sample.
In \cref{fig:Comparisons}, first, we plot a trend line for each data set, both for the exact version, not using our acceleration, and the accelerated version where we approximate the second sum in \cref{equ:KLGradientSplit} via the polar quadtree as described in \cref{sec:AHierarchicalAccelerationStructureForHyperbolicT-SNE}.
Note that \cref{fig:Comparisons}, first, uses a log-scale on both the~$x$- and the~$y$-axis, hence the dashed trend line for~${\mathcal{O}(n^2)}$ becomes a linear graph of slope~$2$.

The exact computation of the gradient (\cref{equ:KLGradient}) on a set of~$n$ points requires~$\mathcal{O}(n^2)$ operations.
Hence, in \cref{fig:Comparisons}, first, we observe a quadratic growth of the average iteration time.
In contrast to that, our accelerated embeddings build a hierarchical data structure, which has a theoretical run time of~$\mathcal{O}(n\log(n))$.
The time taken is mostly two orders of magnitude below the time taken by the exact method, which amounts to a significant speed-up.
This still holds, even when taking variation into account, which can be confirmed in \cref{tab:RunTimeStatistics}, where we report statistics on the run times.
Values are given for the respective full data set.
Note that for all data sets, on average, an iteration saves one---for larger data sets even two---orders of magnitude of run time.

\subsubsection{Estimated Asymptotic Run Time}
\label{sec:EstimatedAsymptoticRunTime}

We will use the experimental data to estimate the asymptotic cost of the computation.
To do so, for a pair of input sizes~$(n_i,n_{i+1})$ and corresponding average iteration times~$(t_i,t_{i+1})$ on these input sizes, we estimate the order~$\alpha$ of the asymptotic run time~${\mathcal{O}(n^{\alpha})}$ as
$\alpha = \frac{\log(t_{i+1}) - \log(t_i)}{\log(n_{i+1}) - \log(n_i)}$,
which is the inverse of the experimental order of convergence, as adapted from Senning~\cite[Equ.~(9)]{senning2007computing}.

As we run ten sample sizes on each data set, we obtain nine estimated values of~$\alpha$ by comparing the run time of each sample size with the run time on the next larger sample.
Consider \cref{fig:Comparisons}, third, for a plot of these estimates.
First, we can observe that the asymptotic order for the exact method is estimated around~${\alpha=2}$, that is, the exact approach takes quadratic run time.
Second, while the estimated asymptotic order fluctuates with the different data sets for our method, the convergence rate~$\alpha$ is always well below~$2$.
This shows that there is an asymptotic time gain, which highlights the increased impact of our method for growing data set sizes.

\begin{figure}
    \centering
    \includegraphics[width=1.\linewidth]{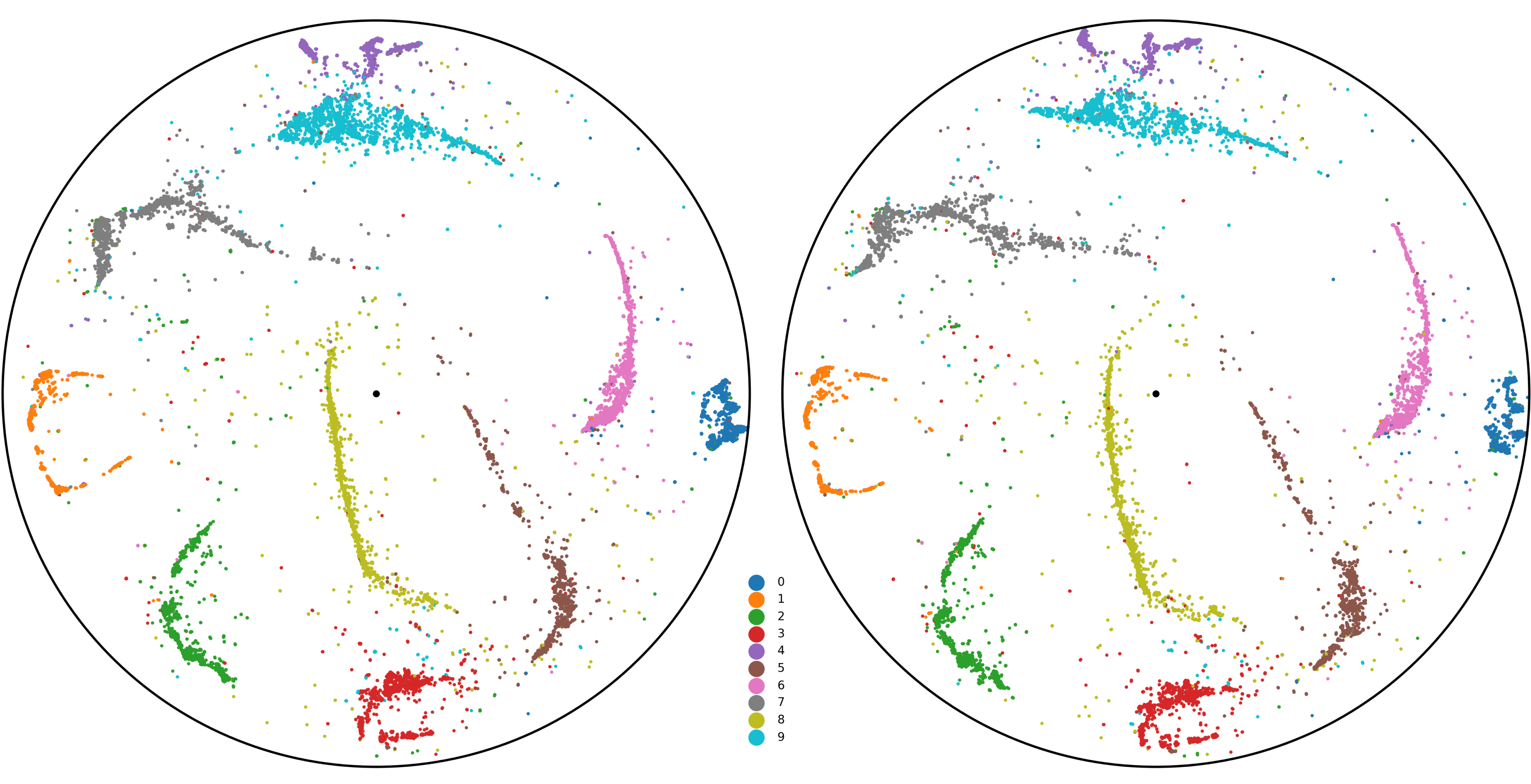}
    \caption{
        Embedding the MNIST data set into the Poincar\'e disk exactly (left) and using our accelerated method (right).
    }
    \label{fig:MNIST}
\end{figure}

\subsubsection{Effect of~$\theta$ on the Run Time}
\label{sec:EffectOfThetaOnTheRunTime}

Just like with the Barnes-Hut data structure~\cite{vanDerMaaten2014accelerating}, the main steering parameter of our acceleration is~$\theta$, which determines whether or not a subtree of the hierarchy is explored or approximated following~\cref{equ:CellApproximationCriterion}.
For~$\theta=0$, no approximation is performed, and for growing values of~$\theta$, an increasing number of subtrees is approximated.
We measure this effect by embedding the full data sets listed in~\cref{tab:Datasets} with varying approximation values~${\theta\in\{0,0.1,\ldots,1.0\}}$, using 250~iterations of early exaggeration and 750~iterations of non-exaggerated gradient descent, possibly stopping earlier, as discussed in \cref{sec:ReductionOfTheAbsoluteRunTime}.
We then report the average run time of these iterations, see~\cref{fig:Comparisons}, fourth.

We see a similar effect as with the Barnes-Hut data structure~\cite{vanDerMaaten2014accelerating}, in the sense that increasing the value of~$\theta$ significantly reduces the run time of the iterations.
When comparing to the behavior of the Barnes-Hut tree in the Euclidean setting~\cite[Fig.~3]{vanDerMaaten2014accelerating}, we see a similar tendency for our method to reach a plateau, which is not as notable with the log-scale y-axis in~\cref{fig:Comparisons}, fourth, as it is with the linear scale of~\cite[Fig.~3]{vanDerMaaten2014accelerating}.
Yet, these experiments show that our method successfully replicates the run time reduction from Euclidean t-SNE, but for hyperbolic t-SNE implementations.


\subsubsection{Time Gain by Splitting Choices}
\label{sec:TimeGainBySplittingChoices}

The previous experiments were run with a polar quadtree built by splitting according to \cref{equ:EqualEmbeddingLengthsSplitting}, as opposed to the splitting strategy proposed in the original publication, see \cref{equ:EqualAreaSplitting}.
To show that this change in the data structure has a positive effect on the run time, we repeat the experiment described in \cref{sec:ReductionOfTheAbsoluteRunTime}, but this time, we compare the two different splitting options, see \cref{fig:Comparisons}, second.

Averaged over all runs, utilizing our proposed splitting option \cref{equ:EqualEmbeddingLengthsSplitting} outperforms the original splitting method \cref{equ:EqualAreaSplitting}.
While this gain is different across the data sets, we obtain an average time reduction per iteration of 11\% for Lukk, 20\% for Myeloid, 64\% for Planaria, 34\% for MNIST, 42\% for WordNet, and 45\% for C.Elegans.
As a concrete example, in the largest size of C.Elegans, with 89,701 points, the average iteration run time goes down from about~12 seconds for equal area splitting (\cref{equ:EqualAreaSplitting}) to roughly~6 seconds for equal lengths splitting (\cref{equ:EqualEmbeddingLengthsSplitting}).
This justifies our choice of implementing a new splitting behavior for the acceleration data structure, which also sets our data structure apart from the polar quadtree as presented in previous work~\cite{looz2015generating}.


\subsection{Quality Retention under Acceleration}

We now turn to the quality of the obtained embeddings.
To measure the embedding quality, on the one hand, we turn to method-intrinsic measures, such as the norm of the gradient field \cref{equ:KLGradientHyperbolic} and the value of the cost function \cref{equ:KLDivergence}.
On the other hand, we investigate method-extrinsic measures, such as the one-nearest neighbor error~\cite{vanDerMaaten2014accelerating}.

\subsubsection{Retention of the Gradient and the Cost Function}

Ultimately, our acceleration method approximates the exact variations (\cref{equ:KlGradientHyperbolicSplit}) by introducing the summarized terms (\cref{equ:CellApproximation}).
To measure the quality of this approximation, we compare the summarized gradient term to the exact one.
Experimentally, for each of the six data sets from \cref{tab:Datasets} and for each of the sample sizes from \cref{sec:ReductionOfTheAbsoluteRunTime}, we compute a hyperbolic embedding using our accelerated method performing~250 iterations of early exaggeration and~750 iterations of non-exaggerated gradient descent.
At iterations~${i\in\{0,50,100,150,200,249\}}$ of early exaggeration and iterations~${i\in\{0,50,\ldots,700,749\}}$ of non-exaggerated gradient descent (possibly less if stopping earlier, see \cref{sec:ReductionOfTheAbsoluteRunTime}), we compute the relative error of the exact version~$\mathbf{g}_i$ and the approximated version~$\hat{\mathbf{g}}_i$ as
$\frac{\sqrt{\sum_{i}d^\mathcal{H}\left(\mathbf{g}_i, \hat{\mathbf{g}}_i\right)^2}}{\sqrt{\sum_{i}d^\mathcal{H}\left(\mathbf{0}, \mathbf{g}_i\right)^2}}$,
that is we relate the norm of the distance field between the two gradients to the norm of the exact gradient field.
In \cref{tab:GradientAndCostFunctionErrors}, we report the average of all such relative errors, measured at the iterations indicated above, for each of the sample sizes and runs as laid out in \cref{sec:ReductionOfTheAbsoluteRunTime}.
From the values presented, we can conclude that the relative approximation error of the gradient in each iteration is about~$1$ to~$2.7$ permille.

\begin{table}[htb]
    \centering
    \caption{
        Mean relative gradient and cost function error.
    }
    \renewcommand*{\arraystretch}{1.2}
    \begin{tabular}{l|c|c}
        Data set             & Gradient           & Cost Function \\
        \hline
        Lukk et al.         & $1.754\cdot10^{-3}$ & $0.343\cdot10^{-6}$\\
        MyeloidProgenitors  & $1.141\cdot10^{-3}$ & $0.801\cdot10^{-6}$\\
        Planaria            & $0.974\cdot10^{-3}$ & $2.357\cdot10^{-6}$\\
        MNIST               & $1.673\cdot10^{-3}$ & $1.690\cdot10^{-6}$\\
        WordNet             & $2.715\cdot10^{-3}$ & $0.654\cdot10^{-6}$\\
        C.Elegans           & $2.015\cdot10^{-3}$ & $0.343\cdot10^{-6}$
    \end{tabular}
    \label{tab:GradientAndCostFunctionErrors}
\end{table}

\begin{figure*}
    \centering
    \includegraphics[width=1.\textwidth]{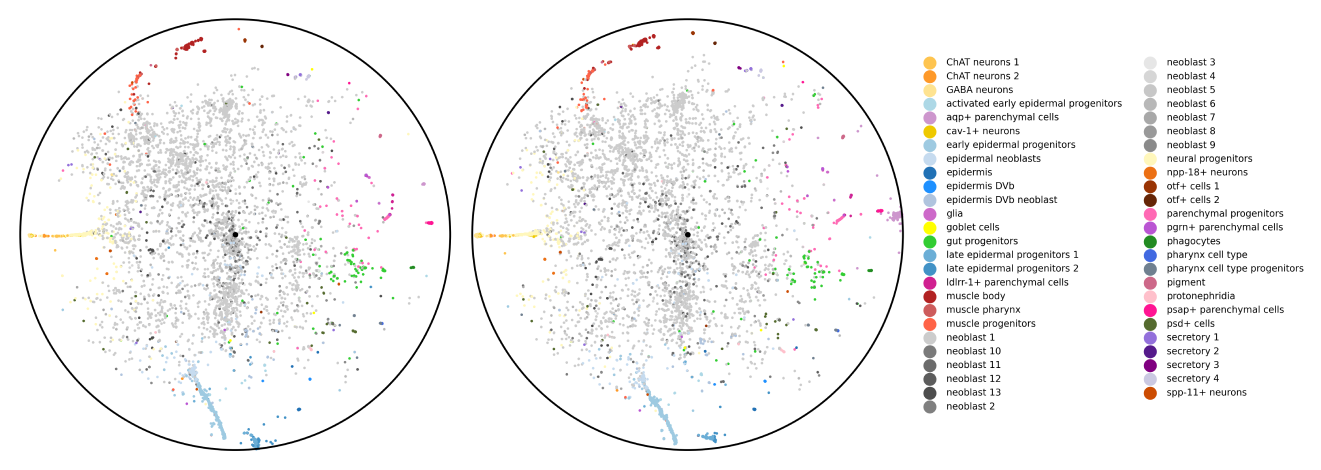}
    \caption{
        Embedding the Planaria data set into the Poincar\'e disk exactly (left) and using our accelerated method (right). 
    }
    \label{fig:Planaria}
\end{figure*}

Furthermore, we turn to the cost function of the full embedding obtained after all iterations.
We evaluate the cost function (\cref{equ:KLDivergence}), utilizing the hyperbolic low-dimensional probabilities (\cref{equ:LowDimensionalPorbabilityHyperbolic}), providing an exact value~$C$ of the non-accelerated embedding and a cost function value~$C'$ of the accelerated embedding.
We compute the relative error of these as~${|C-C'|/C}$.
In \cref{tab:GradientAndCostFunctionErrors}, we present the mean of all these relative cost function errors across the data sets, averaged over the runs explained in \cref{sec:ReductionOfTheAbsoluteRunTime}.
This shows that while there is a gradient approximation error of about~$1$ to $2.7\textperthousand$, the effect on the cost function of the final embedding is three orders of magnitude smaller.
Hence, our method is efficient at accelerating the embedding procedure while not affecting the quality of the results, see the qualitative comparison in \cref{fig:MNIST} and \cref{fig:Planaria}.

\subsubsection{Effect of the Acceleration on the Embedding Quality}
\label{sec:EffectOfTheAccelerationOnTheEmbeddingQuality}

It was shown for the Barnes-Hut acceleration method, that larger values of~$\theta$ lead to larger 1-nearest neighbor errors in corresponding embeddings~\cite[Fig.~3]{vanDerMaaten2014accelerating}.
The 1-nearest neighbor error is given by the percentage of points whose nearest neighbor in the embedding does not have the same class label as the query point.
Note that for~${k=1}$, this is the inverse of the neighborhood hit, as discussed by Espadoto et al.~\cite{espadoto2019toward}.
We utilize this measure to investigate the effect of the acceleration on the embedding quality.
See \cref{tab:OneNearestNeighborErrors} for the errors obtained on five of the six data sets.
Note that the WordNet data set has been removed from this measure as it does not have a clear cluster structure.

\begin{table}[htb]
    \centering
    \caption{
        The 1-nearest neighbor error for the exact and accelerated embeddings with~$\theta=0.5$.
    }
    \renewcommand*{\arraystretch}{1.2}
    \begin{tabular}{c|r|r}
        Data Set            & Exact     & Accelerated \\
        \hline
        Lukk et al.         &  4.13\%   &  4.11\% \\
        MyeloidProgenitors  & 20.46\%   & 21.39\% \\
        Planaria            & 15.89\%   & 16.12\% \\
        MNIST               &  4.12\%   &  3.96\% \\
        C.Elegans           & 29.40\%   & 26.07\% \\
    \end{tabular}
    \label{tab:OneNearestNeighborErrors}
\end{table}

The experimental results in \cref{tab:OneNearestNeighborErrors} show that the local quality of the embedding, measured by the 1-nearest-neighbor error remains the same under acceleration.
That is, our method produces embeddings that capture local cluster structures, as well as the exact hyperbolic t-SNE formulation while being considerably faster.

To further quantify the quality of our embeddings, we turn to the precision/recall metric~\cite{pezzotti2016hierarchical}.
For that, we follow the previous work and fix a maximum neighborhood size~${k_{\max}=30}$.
Then, for each~${k\in\{1,\ldots,k_{\max}\}}$, we compute the number of true positives as $\TP_k=N_{k_{\max}}(X)\cap N_K(Y)$, that is, the points that are in the high-dimensional neighborhood and also in the low-dimensional embedded neighborhood, given the respective metrics.
From this value, we obtain the precision as
$\PR_k=|\TP_k|/k$ and the recall as $\RC_k=|\TP_k|/k_{\max}.$
That is, ideally, the precision is always~$1$, while the recall grows as~${k/k_{\max}}$, yet, a data set might not exhibit such a solution, nor does t-SNE necessarily find this solution.
Instead, we want to show that our acceleration does not influence this resulting quality significantly while achieving a significant speedup.

\begin{figure*}
    \newlength{\NNPWidth}
    \setlength{\NNPWidth}{1.0\textwidth}
    \includegraphics[width=\NNPWidth]{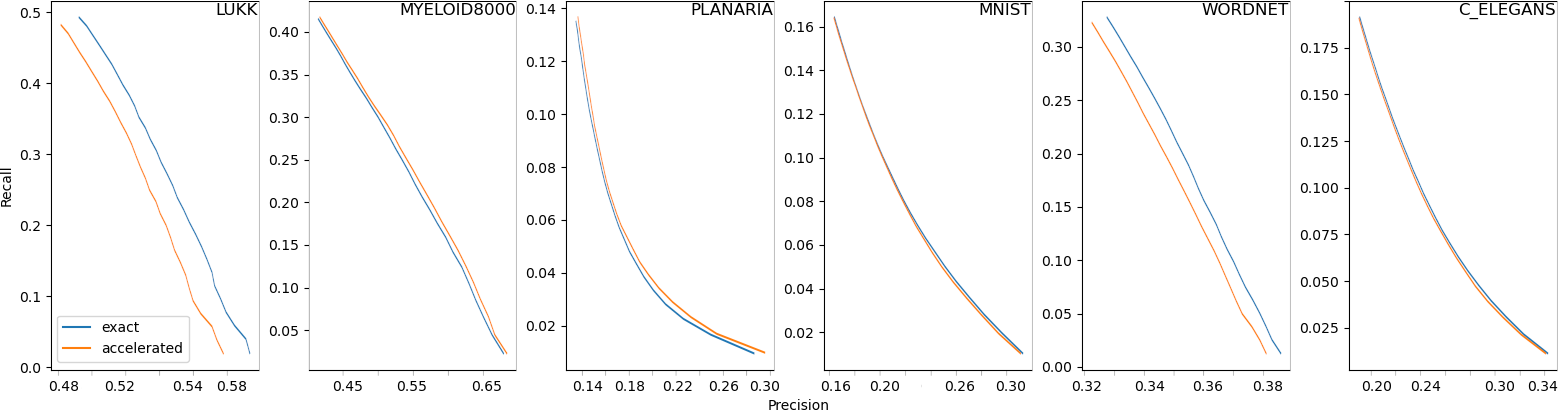}
    \caption{Comparison of the precision/recall curves for exact and accelerated hyperbolic embeddings.}
    \label{fig:PrecisionRecallCurves}
\end{figure*}

See precision/recall curves for all data sets in Figure~\ref{fig:PrecisionRecallCurves}.
While the preservation of the accelerated embeddings falls slightly off, in particular for the larger embeddings, overall the preservation behavior is similar.
This further demonstrates how our acceleration keeps local neighborhoods at a quality comparable to that of the exact method while obtaining the embeddings significantly faster.

\subsubsection{Effect of~$\theta$ on the Embedding Quality}

\begin{figure}[h!]
    \centering
    \includegraphics[width=.9\linewidth]{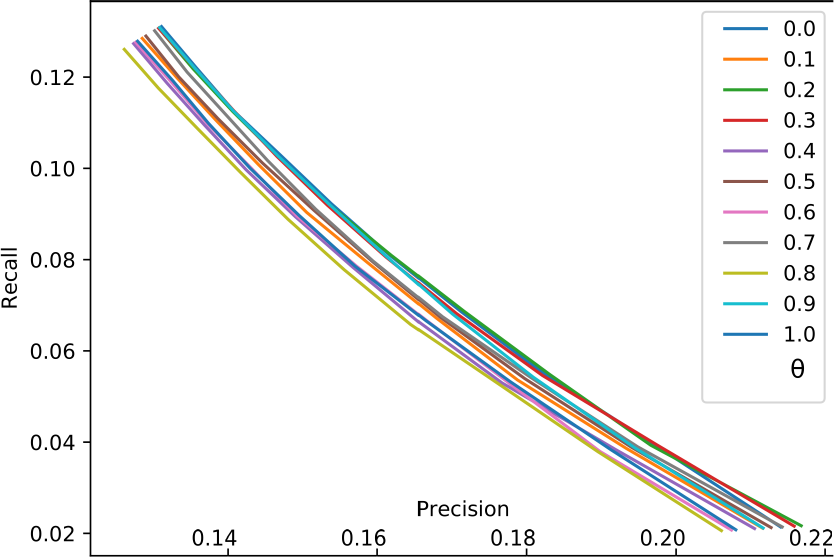}
    \caption{Nearest-neighbor preservation on the MNIST data set for varying values of~$\theta$.}
    \label{fig:PrecisionRecallMNISTvaryingTheta}
\end{figure}

As discussed in \cref{sec:EffectOfThetaOnTheRunTime}, the main parameter of our acceleration data structure is~$\theta$.
Here, we investigate the effect different choices of~$\theta$ have on the quality of the final embedding.
We choose~${\theta\in\{0.0,0.1,0.2,\ldots,1.0\}}$ and embed the MNIST data set using 250 iterations of early exaggeration and up to 750 iterations of non-exaggerated gradient descent, possibly stopping earlier, as discussed in \cref{sec:ReductionOfTheAbsoluteRunTime}.
For each of the embeddings, we then compute the neighborhood preservation via a precision/recall curve.
These curves are shown in \cref{fig:PrecisionRecallMNISTvaryingTheta}.
While the approximations fall off slightly when compared to the exact solution~${\theta=0.0}$, the neighborhood quality remains very stable within the explored range of~$\theta$.

Note that~${\theta=0}$ is the exact version and obtains the best, upper-rightmost curve.
However, the curves for other, increasing values of~$\theta$ are very comparable and do not fall off significantly in comparison.
Therefore, we deduce that the acceleration has a small enough effect on the embedding quality to make its time gains a reasonable trade-off.


\subsection{Embedding of Large-Scale Data Sets}

\begin{figure}
    \includegraphics[width=1.\linewidth]{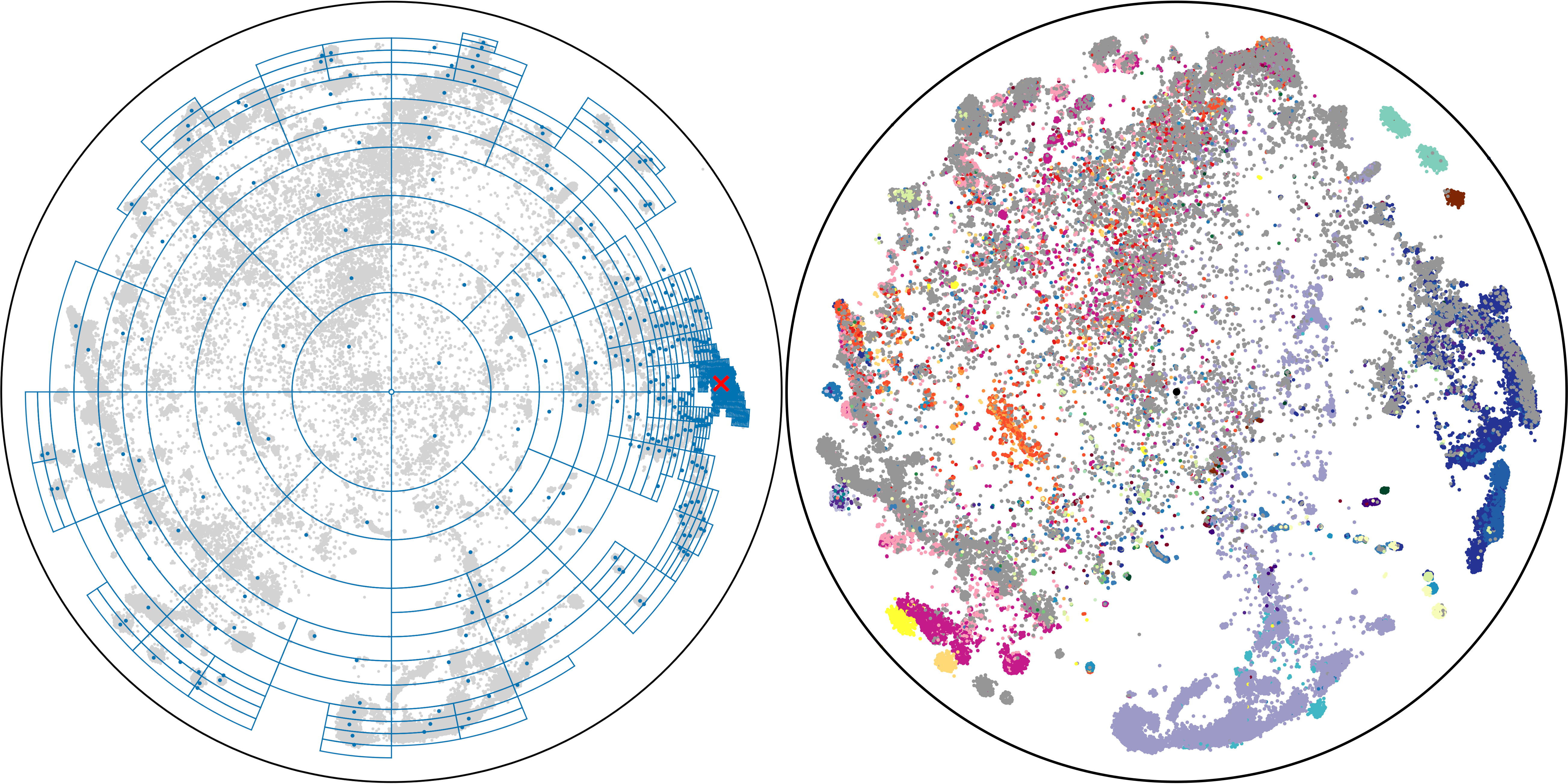}
    \label{fig:teaser}
    \caption{Left: Our polar quadtree acceleration structure on top of the C.Elegans data points.
        The red mark indicates a query point and the polar quads include groups of points that are summarized via the quad midpoints, shown as blue dots.
        Right: Final embedding of the full C.Elegans data set~(89,701 points) into hyperbolic space.}
\end{figure}

Due to the slow processing of larger data sets, previous approaches turned to subsampling data to show hyperbolic embeddings.
For instance, the Poincar\'e maps approach took a sub-sample of~40,000 data points from the C.Elegans data set, and its GPU implementation spent~2--3 hours to compute an embedding~\cite{klimovskaia2020poincare}.
With our acceleration structure, we can embed not only the full C.Elegans data set (\cref{fig:teaser}), but also to do so on the CPU within~45~minutes.
Generally, these time gains mean that, with our acceleration structure, hyperbolic embeddings can be computed without costly graphics cards and thus become more widely available to researchers.
The larger size of data sets that can be handled furthermore unlocks previously infeasible application scenarios.

\section{Conclusion}

In this paper, we have presented an acceleration data structure for hyperbolic t-SNE embeddings and discussed how to approximate the hyperbolic gradient with it.
We have shown that this approach is a potential building block for existing and upcoming hyperbolic embedding techniques.
Our experimental results validate the time gain while showing that there is no significant loss in embedding quality.

Our work focuses on hyperbolic t-SNE-type embeddings, and extending our acceleration to other embedding approaches is left for future work.
Furthermore, it remains to be investigated how to translate Fourier transform approaches or stochastic gradient descent to the hyperbolic t-SNE scenario.
While there are generalizations of Fourier transform to hyperbolic space~\cite{isozaki2014fourier,xiao2022complex}, the main challenge for the context of embedding computations would be to build a regular grid with a clear control on the number of grid points and the grid-cell shape in hyperbolic space.
As for stochastic gradient descent, sampling the gradient causes the repulsive forces of the cost function to become unbalanced with the attractive forces.
The main challenge lies in balancing the sampling rate with a re-normalization of the forces.
Furthermore, it remains to be investigated how these approximations affect both the run time and the embedding quality.


\section*{Supplemental Materials}
\label{sec:supplemental_materials}

The data sets used in our experiments are available as follows: Lukk et al.~(\url{https://www.ebi.ac.uk/biostudies/arrayexpress/studies/E-MTAB-62}), MyeloidProgenitors~(\url{https://github.com/scverse/scanpy_usage/tree/master/170430_krumsiek11}), Planaria~(\url{https://shiny.mdc-berlin.de/psca/}), MNIST~(\url{https://yann.lecun.com/exdb/mnist/}), C.Elegans~(\url{https://github.com/Munfred/wormcells-data/releases}), and WordNet~(\url{https://github.com/facebookresearch/poincare-embeddings}).

The implementation of the acceleration structure as well as all scripts for the experiments outlined in this paper can be found at \url{https://graphics.tudelft.nl/accelerating-hyperbolic-tsne}.

\section*{Figure Credits}
\label{sec:figure_credits}

\Cref{fig:BarnesHutApproximation} is an adapted reprint of~\cite[Fig.~2]{vanDerMaaten2014accelerating}.
\Cref{fig:PoincareDisk}, left, is a reprint of \url{https://commons.wikimedia.org/wiki/File:Poincare_disc_hyperbolic_parallel_lines.svg}, which is in the public domain.
\Cref{fig:PoincareDisk}, right, is a reprint of \url{https://commons.wikimedia.org/wiki/File:*732_tiling_on_the_Poincar%C3%A9_Disk.svg}, by Murdock Grewar, available under the Creative Commons Attribution-Share Alike 4.0 International license.

\section*{Acknowledgments}%
Martin Skrodzki's work was funded by the Deutsche Forschungsgemeinschaft (DFG, German Research Foundation) -- 455095046. Nicolas F.\@ Chaves-de-Plaza was funded in part by Varian, a Siemens Healthineers Company, through the HollandPTC-Varian Consortium under Grant 2019022, and in part by the Surcharge for Top Consortia for Knowledge and Innovation (TKIs) from the Ministry of Economic Affairs and Climate.

\bibliographystyle{abbrv-doi-hyperref}
\bibliography{literature}

\section{Biographies}

\begin{IEEEbiography}[{\includegraphics[width=1in,height=1.25in,clip,keepaspectratio]{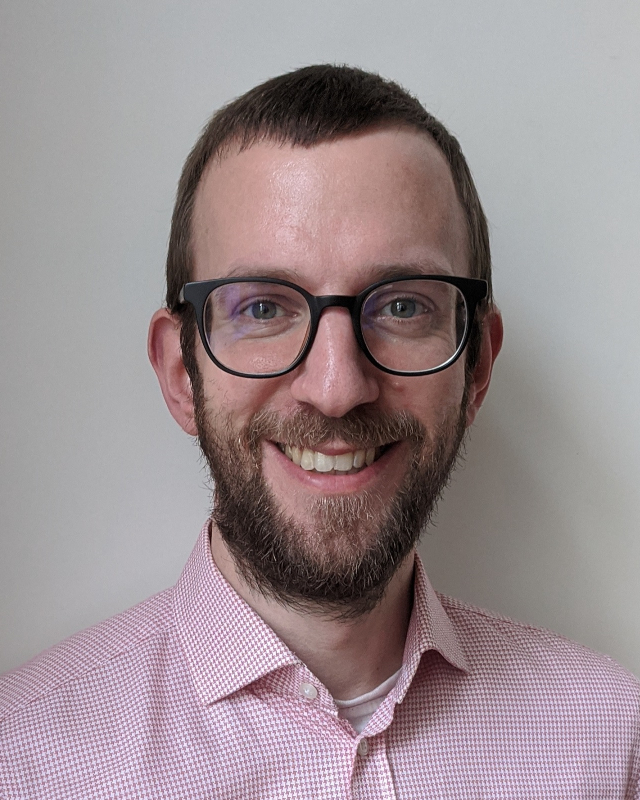}}]{Martin Skrodzki} is an assistant professor with a special focus on education at the Computer Graphics and Visualization Group at the Department of Intelligent Systems at Delft University of Technology.
His research interests include the use of illustrations in mathematics, visualization of high-dimensional data, discrete geometry processing as well as interactions between mathematics and arts.
\end{IEEEbiography}

\begin{IEEEbiography}[{\includegraphics[width=1in,height=1.25in,clip,keepaspectratio]{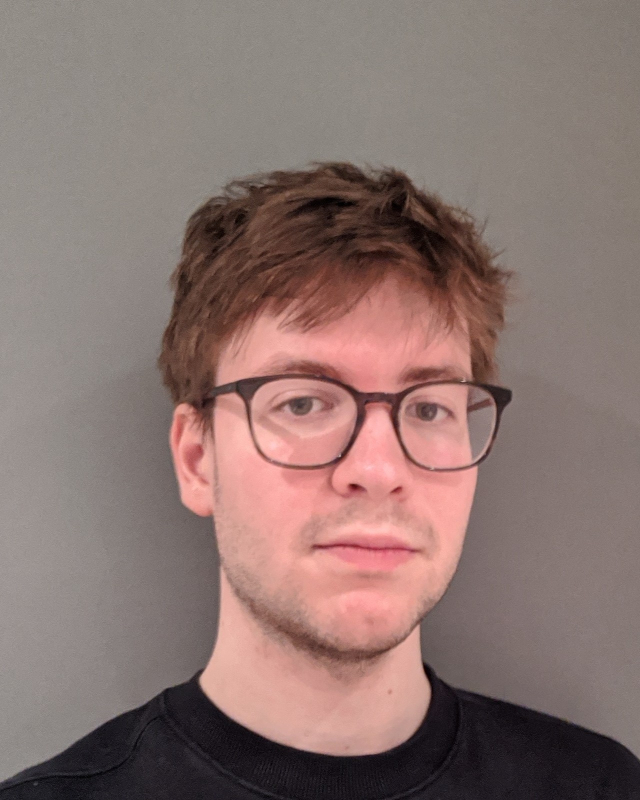}}]{Hunter van Geffen} studied Computer Science at the Delft University of Technology where he graduated with a Master's Degree in 2022.
His studies focused on Data Science and Visualization and he wrote his thesis about accelerating T-SNE for hyperbolic embeddings.
Currently, he is a Scientific Software Engineer at VORTech.
\end{IEEEbiography}


\begin{IEEEbiography}[{\includegraphics[width=1in,height=1.25in,clip,keepaspectratio]{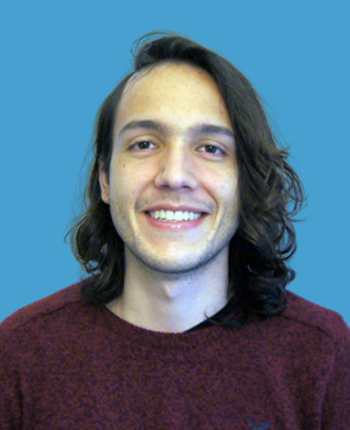}}]{Nicolas F.\@ Chaves-de-Plaza} is a Ph.D. student at the Department of Intelligent Systems at the Delft University of Technology. He works on developing visualization and interaction tools to support clinician-driven segmentation of 3D medical images in the context of Adaptive Proton Therapy.
\end{IEEEbiography}


\begin{IEEEbiography}[{\includegraphics[width=1in,height=1.25in,clip,keepaspectratio]{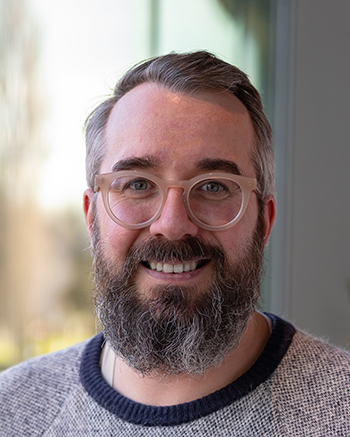}}]{Thomas H\"ollt} is a tenured Assistant Professor at the Computer Graphics and Visualization Group at the Department of Intelligent Systems at Delft University of Technology. His research interests include Visualization and Visual Analytics, with a focus on high-dimensional data and bio-/medical applications.
\end{IEEEbiography}


\begin{IEEEbiography}[{\includegraphics[width=1in,height=1.25in,clip,keepaspectratio]{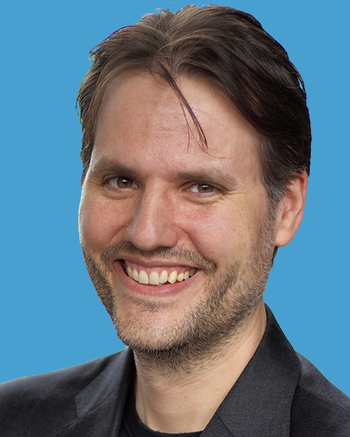}}]{Elmar Eisemann} is a Full Professor heading the Computer Graphics and Visualization Group at the Department of Intelligent Systems at Delft University of Technology. His research interests include real-time and perceptual rendering, visualization, alternative representations, shadow algorithms, global illumination, and GPU acceleration techniques.
\end{IEEEbiography}


\begin{IEEEbiography}[{\includegraphics[width=1in,height=1.25in,clip,keepaspectratio]{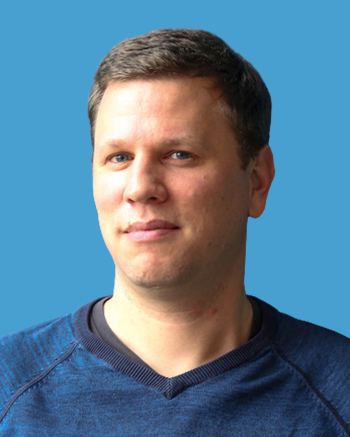}}]{Klaus Hildebrandt} is an Assistant Professor at the Computer Graphics and Visualization Group at the Department of Intelligent Systems at Delft University of Technology. His research interests include Visual Computing, Geometric Data Processing, Physical Simulation, and Computational and Discrete Differential Geometry.
\end{IEEEbiography}

\end{document}